\documentclass[preprint,authoryear,12pt]{elsarticle}
\usepackage{amssymb}
\usepackage{color}
\usepackage{amsmath}
\usepackage{subfigure}
\usepackage{graphicx}
\usepackage{graphics}
\usepackage{booktabs}
\usepackage{threeparttable}
\usepackage{appendix}
\usepackage{geometry}

\usepackage[displaymath]{lineno}
\usepackage{setspace}
\usepackage{soul}
\usepackage{lscape}
\usepackage{multirow}
\usepackage{caption}
\DeclareCaptionLabelFormat{cont}{#1~#2\alph{ContinuedFloat}}
\usepackage{hyperref}

\usepackage{mathpazo}

\journal{ }

\begin{document}

\begin{frontmatter}

\title{Investigating and modeling day-to-day route choices based on laboratory experiments. 
Part II: A route-dependent attraction-based stochastic process model}

\author[rvt1,rvt3]{Hang Qi\corref{firstauthor}}

\author[rvt2,rvt3]{Ning Jia\corref{firstauthor}}

\author[rvt4]{Xiaobo Qu}

\author[rvt5]{Zhengbing He\corref{cor1}}

\ead{he.zb@hotmail.com}

\address[rvt1]{Institute for Advanced Studies in Finance and Economics, Hubei University of Economics, China}
\address[rvt2]{Institute of Systems Engineering, College of Management and Economics, Tianjin University, China}
\address[rvt3]{Laboratory of Computation and Analytics of Complex Management Systems (Tianjin University), China}
\address[rvt4]{State Key Laboratory of Automotive Safety and Energy, Tsinghua University, China}
\address[rvt5]{Beijing Key Laboratory of Traffic Engineering, Beijing University of Technology, China}

\cortext[firstauthor]{Co-first authors}

\cortext[cor1]{Corresponding author}

\begin{spacing}{1.2}

\begin{abstract}
To explain day-to-day (DTD)  route-choice behaviors and traffic dynamics observed in a series of lab experiments, Part I of this research proposed a discrete choice-based analytical dynamic model \citep{QI2023103553}. Although the deterministic model could well reproduce the experimental observations, it converges to a stable equilibrium of route flow while the observed DTD evolution is apparently with random oscillations. To overcome the limitation, the paper proposes a route-dependent attraction-based stochastic process (RDAB-SP) model based on the same behavioral assumptions in Part I of this research. Through careful comparison between the model-based estimation and experimental observations, it is demonstrated that the proposed RDAB-SP model can accurately reproduce the random oscillations both in terms of flow switching and route flow evolution. To the best of our knowledge, this is the first attempt to explain and model experimental observations by using stochastic process DTD models, and it is interesting to find that the seemingly unanticipated phenomena (i.e., random route switching behavior) is actually dominated by simple rules, i.e., independent and probability-based route-choice behavior. Finally, an approximated model is developed to help simulate the stochastic process and evaluate the equilibrium distribution in a simple and efficient manner, making the proposed model a useful and practical tool in transportation policy design. 

\end{abstract}
\end{spacing}


\end{frontmatter}




\newpage

\begin{spacing}{1.5}
\section{Introduction}

Day-to-day (DTD) traffic dynamics has attracted much attention to describe the transient behavior of network flow from non-equilibrium states to equilibrium states.
Theoretical models for DTD traffic dynamics are built either on travelers' route choice behavior or flow swapping among routes. 
Experiment-based research is an important stream to bridge the gap between theoretical and real-world studies on DTD traffic dynamics, whereas the latter are limited by the difficulty of collecting network-wide high-quality route choice-related data \citep{Guo2011a,Zhu2011,He2012b,watling2012network}.

In Part I of this research, we conducted a series of multi-scenario multi-repetition DTD route-choice experiments and proposed a discrete choice-based DTD model to explain the newly-observed phenomena \citep{QI2023103553}.
It was shown that the proposed model could satisfactorily reproduce the experimental observations such as the switching rates and equilibrium flow.
However, as with many other differential function-based models \citep{Watling1999, He2012b, Cantarella2016,He2016b, Kumar2015, Xiao2016, Ye2017, xiao2019day-to-day},
the proposed model converges to a stable equilibrium prediction of route flow, which deviates from the experimental observations of our experiments and those experiments in the existing studies \citep{Iida1992,Selten2007a,Zhao2016c,Ye2018,Qi2019,Han2021}. 
The primary reason is that those models are all deterministic, in which the network flow evolution processes are fully determined (with the probability of 1) given the initial conditions.

The stochastic process DTD model \citep{cascetta1989a,cantarella1995dynamic} is a potential means to fill this gap. In the stochastic process models, the current flow assignment only determines the distribution of flow assignment in the next time, rather than a one-on-one projection as in the deterministic models. 
Also, the assignment converges 
to its equilibrium probability distribution of the flow states instead of a single steady-state equilibrium, which may vary according to the behavioural and traffic-related assumptions made in the stochastic process. 
Therefore, the stochastic process models provide a description of not only 'expected' dynamic behaviour but also probabilistic oscillations in the network flow evolution process.

A handful of stochastic process DTD models have been proposed.
\cite{cascetta1989a} may be the one that first proposed a stochastic process model.
It pointed out that the sequence of network states over successive times is the realization of a stochastic process, as no transportation system remains in the same state all the time. 
Later, \cite{Davis1993Large} proposed two general stochastic process DTD models and demonstrated that the general models can be approximately regarded as a deterministic process or a time-varying linear Gaussian process. 
\cite{cantarella1995dynamic} attempted to unify the deterministic and stochastic processes of DTD dynamics modelling of network flows. 
With large enough travel demand and link capacity, the gap between deterministic and stochastic descriptions of system evolution would be shrunk. 
\cite{watling2013modelling} analyzed different kinds of stochastic process models, and showed the great potential of stochastic process models to capture various contributory sources of variance in real transportation systems.
To better describe the real-life uncertainty, \cite{smith2014long} proposed two approaches to introduce randomness into DTD dynamics. 
The first approach is to embed a stochastic element into a deterministic process, and the second one is based on the independent and stochastic individual traveller's choice.
More recently, \cite{Cantarella2016} proposed a general stochastic process DTD dynamic model by considering traveler's habit and memory over past days.



As shown in the experiments in Part I, the network flow evolution has a non-deterministic nature and it cannot be well reproduced by deterministic models \citep{QI2023103553}.
Therefore, it is interesting to see if a stochastic model could reflect the non-deterministic nature. 
However, to the best of our knowledge, no previous research has been carried out by using the stochastic DTD model to explain experimental observations. 
The reason might be two-fold. 
First, the data in most laboratory experiments is not rich enough for exploring statistic laws. 
Second, no existing models could reproduce experimental observations satisfactorily. 
Given that much richer data have been obtained and several key behavioral rules have been extracted in Part I of the research \citep{QI2023103553}, we could overcome the two obstacles above-mentioned. 
This part of the research investigates experimental DTD dynamics by proposing a route-dependent attraction-based stochastic process (RDAB-SP) model based on exactly the same behavioral assumptions in Part I \citep{QI2023103553}. 
In addition, stochastic process-based models usually result in high computational burden, which hinders the model from practical usage. 
To overcome the drawback, we further propose an approximate model that can simplify the simulation process with acceptable loss of accuracy.
%
%
The contributions of this paper are as follows.
\vspace{-0.3cm}
\begin{itemize}
\setlength{\itemsep}{0pt}
\setlength{\parsep}{0pt}
\setlength{\parskip}{0pt}	

	\item \textcolor{black}{A new stochastic process model, named the RDAB-SP model, is proposed. 
    It can accurately reproduce (1) the random oscillations in not only route flows but also route switching behavior observed in the DTD route-choice experiments; (2) the stochastic characteristics of the route flow evolution process including the switching oscillation and the route flow oscillation, indicating that the unanticipated behavior regularities observed in laboratory experiments (i.e., route-dependent inertia and preference) are actually dominated by simple rules, i.e., independent and probability-based route choice.    
    }
	
	
	\item An approximation of the RDAB-SP model is developed, which helps to simulate the stochastic process and evaluate the equilibrium distribution in a simple and efficient manner.
	The simple version will effectively facilitate the model to be a useful and practical tool for transportation planners and policy makers.
		
\end{itemize}

To the best of our knowledge, this study is the first attempt to explain and model experimental observations by using stochastic process models, \textcolor{black}{and also the first one that successfully reproduces the random oscillations of route flows.}
It confirms the generality of the route-choice behavior framework proposed in Part I of the research, i.e., the discrete choice-based model with route-dependent inertia and preference \citep{QI2023103553}. 
It also marks the advantage of explicit modeling of individual choice behavior compared to modeling of flow swapping rules, as the proposed RDAB-SP model is constructed based on individual's independent choice.

Moreover, the proposed stochastic model would be much more appropriate for analyzing the impact of transportation policies, resulted from the advantage of the proposed model in capturing the probability distribution of long-term network impacts, not just the expected trend or equilibrium points. 
Therefore, it would greatly help transportation engineers and planners to understand and improve the robustness of a transportation policy.

This paper is organized as follows.
Section \ref{sec:Background} briefly introduces the research conducted in Part I \citep{QI2023103553}, including the DTD route-choice experiments and the deterministic discrete choice-based model. 
Section \ref{sec:Model} proposes a RDAB-SP model and Section \ref{sec:Test} carefully tests it by using the experimental data.
Section \ref{sec:Approximate} proposes an approximation of the RDAB-SP model, which helps simulate the stochastic evolution process in a more efficient manner.
Section \ref{sec:Conclusion} concludes the paper.

\section{Background}\label{sec:Background}

\subsection{Day-to-day Route-choice Experiments}

We first briefly introduce the DTD route-choice experiments that were carried out in controlled laboratory environment. \textcolor{black}{One may refer to Part I of the research for more details \citep{QI2023103553}.}

Total 312 participants in approximately equal proportions of males and females were invited to participate the decision-making experiments for the payoffs that were contingent on their performance.
The participants were randomly assigned to 17 groups and they were required to make DTD route choices with no mutual communication.
The following eight DTD scenarios with the same origin-destination (OD) pair were set.
\vspace{-0.3cm}
\begin{itemize}
\setlength{\itemsep}{0pt}
\setlength{\parsep}{0pt}
\setlength{\parskip}{0pt}	  
	\item Scenario 1 was the baseline scenario containing a symmetric two-route network. 
	
    \item Scenarios 2-5 extended Scenario 1 by using asymmetric two-route networks and different cost functions to investigate subject’s route choice behaviors under different cost feedback.
	
	\item Scenarios 6-7 employed asymmetric networks with three routes to observe more complicated route choice behavior.
	
	\item \textcolor{black}{Scenario 8 extended the configuration to \textit{non-linear} cost functions and different group sizes (24 subjects per group in Scenario 8, while 16 in others), which would demonstrate the robustness of the proposed theoretical model.}
	
\end{itemize}
\vspace{-0.3cm}

Those routes were designed to be susceptible to congestion.
Total 16/24 participants were assigned into each scenario and they were instructed that the travel time of a route was an increasing function of the route flow.
After every subject made their route choices, the complete feedback information was then displayed to all participants.
The number of the route-choice rounds was determined by randomly selecting a number between 75 and 110.



Interesting phenomena were observed from the experimental results, including the continuous oscillations in route flow without a steady endpoint, and two behavior patterns which cannot be well explained by the existing models.

\subsection{Deterministic Discrete Choice-based Model}

In Part I of this research, we proposed a discrete choice-based dynamic model written as follows \citep{QI2023103553}.
\begin{equation}\label{equ:hit+1}
    f_j^{t+1} = \sum_{i=1}^N \Delta f_{ij}^t = \sum_{i=1}^N p_{ij}^t f_i^t
\end{equation}
where $f_j^t$ is the number of the travelers on route $j$ at time $t$ (i.e., route flow); 
$\Delta f_{ij}^t$ is the expected flow switching from route $i$ to route $j$ at time $t$; 
$N$ is the size of the feasible set of the routes connecting the same OD pair;
$p^t_{ij}$ is the switching rate from route $i$ to route $j$ at time $t$, and it is defined as the proportion of travelers who switch from route $i$ to route $j$ during time $t$ and $t+1$ ($t \geq 1$). 
The switching rate $p^t_{ij}$ is proposed as follows.
\begin{equation}\label{equ:pji_final}
    p^t_{ij}=
    \begin{cases}
    \begin{split}
    &P_i \frac{e^{-\theta C_j^t}}{\sum_{k=1}^N e^{-\theta C_k^t}}\ , \ & \ \text{if}\ \ i\neq j\\
    &(1-P_i) + P_i\frac{e^{-\theta C_j^t}}{\sum_{k=1}^N e^{-\theta C_k^t}}\ , \ & \ \text{if}\ \ i=j
    \end{split}
    \end{cases}
\end{equation}

Assume that every available route has a route attribute-related attraction coefficient (denoted by $\eta_i$), and the two behavioral patterns, i.e. route-dependent inertia and route-dependent preference, are introduced and incorporated in Equation \ref{equ:pji_final} as follows.

\begin{itemize}
\vspace{-3 mm}
    \setlength{\itemsep}{0pt}
    \setlength{\parsep}{0pt}
    \setlength{\parskip}{0pt}

\item
{\it Route-dependent inertia}. 
Travelers have a constant tendency to remain on their last-chosen route regardless of the cost of the route in the last round. 
The inertia here is route-dependent, indicating that the tendency for travelers varies with different selected routes. 
In Equation \ref{equ:pji_final}, the route-dependent inertia is introduced by setting $P_i=P(\eta_i)$.

\item
{\it Route-dependent preference}. 
The travelers who intend to break the inertia will re-consider their route choices.
The switching decision is affected by not only the cost difference between their last-chosen and alternative routes (commonly considered in the existing models), but also traveler's route attribute-related preference (observed in the experimental data but rarely considered in the existing models). In Equation \ref{equ:pji_final}, the route-dependent preference is incorporated by setting $C_i^t=C(c_i^t, \eta_i)$.
\end{itemize}

\textcolor{black}{The parameters of the model were estimated and the results show that the model performs satisfactorily in terms of Bayesian Information Criterion (BIC), Mean Absolute Percentage Error (MAPE) of average switching rates and equilibrium flow (Table \ref{tab:calibration}).
}

\begin{table}[!htbp]\centering\footnotesize
\setlength{\tabcolsep}{4mm}{
\caption{Calibration results of the proposed deterministic model in the eight scenarios.}\label{tab:calibration}
\begin{tabular}{clllllll}
\toprule
Scenario & $\theta$ & $\eta_1$ & $\eta_2$ & $\eta_3$ & $\text{MAPE}_p$$^{\dagger}$ & $\text{MAPE}_f$$^{\ddag}$ & BIC\\
\midrule
1 &  0.0618 & 0.332 & 0.323 & --- & 0.0896 & 0.0001 & 5647.002 \vspace{2mm}\\
2 &  0.0635 & 0.512 & 0.394 & --- & 0.135 & 0.004 & 4845.526 \vspace{2mm}\\
3 &  0.083 & 0.49 & 0.305 & --- & 0.112 & 0.0058 & 4854.039 \vspace{2mm}\\
4 &  0.019 & 0.448 & 0.192 & --- & 0.0389 & 0.059 & 5980.376 \vspace{2mm}\\
5 &  0.036 & 0.470 & 0.224 & --- & 0.0571 & 0.007 & 6975.715 \vspace{2mm}\\
6 &  0.043 & 0.487 & 0.362 & 0.25 & 0.197 & 0.0419 & 13936.454 \vspace{2mm}\\
7 &  0.0625 & 0.418 & 0.239 & 0.172 & 0.185 & 0.0504 & 14577.203 \vspace{2mm}\\
8 &  0.00875& 0.516 & 0.319 & 0.116 & 0.288 & 0.0784 & 18192.628 \\
\bottomrule
\end{tabular}}
 \begin{tablenotes}\footnotesize
\item $\dagger$ MAPE of the switching rates.
\item $\ddag$ MAPE of the equilibrium flow.
\end{tablenotes}
\end{table}



\subsection{Limitations of Deterministic Model}

Although satisfactory performance has been achieved, the proposed model is still limited by its deterministic nature. 
Specifically, let $\vec f^{t}=(f_1^{t}, f_2^{t}, \cdots, f_N^{t})$ be the network flow assignment at time $t$, and $\vec f^{t+1}$ is only determined by $\vec f^{t}$ in the model.
In contrast, $\vec f^{t}$ could be followed by different flow assignments in the experiments (see Figure \ref{fig:Limitation} for an illustration). 
Hereafter, we call the one-step projection from $\vec f^{t}$ to $\vec f^{t+1}$ as {\it one-step evolution}, which is essentially the conditional distribution of $\vec f^{t+1}$ given $\vec f^{t}$.

The deterministic one-step evolution results in a deviation between the theoretical and experimental flow evolution trajectories.
Taking Scenario 4 where the least estimation error is achieved as an example (Figure \ref{fig:Limitation}), 
it can be seen that the proposed deterministic model produces a smooth trajectory with a stable endpoint, 
while the experimental route flow keeps oscillating with no stable endpoint.
Similar observations are also reported in the existing studies, such as \cite{Iida1992}, \cite{Selten2007a}, \cite{Meneguzzer2013}, \cite{Zhao2016c}, and \cite{Ye2018}.

\begin{figure}[!htbp]
    \centering
    \subfigure[One-step evolution ($\vec f^{t}=(10,6)$)]{
    \includegraphics[width=2.8in]{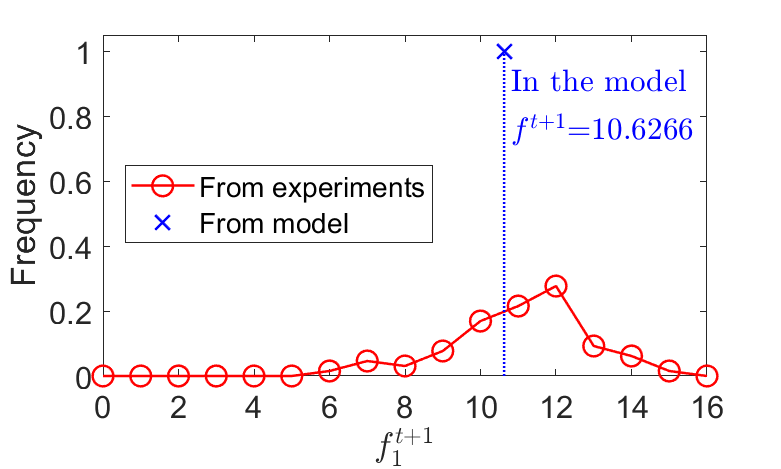}}
    \subfigure[Route flow]{
    \includegraphics[width=2.8in]{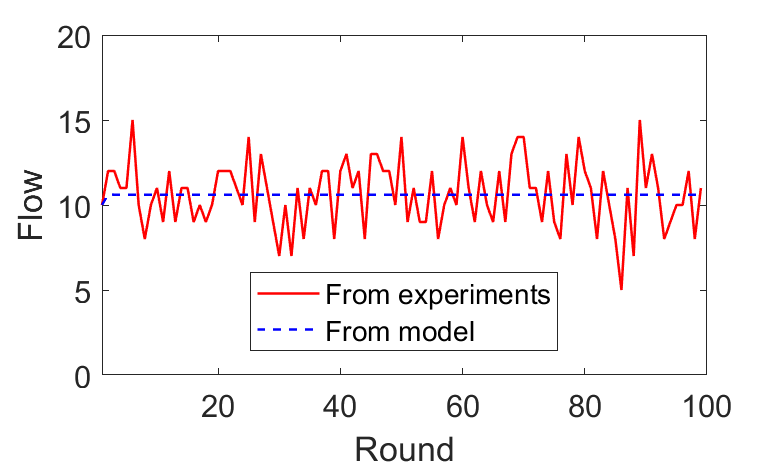}}
    \caption{Limitation of the deterministic model: an illustration by taking the route 1 in Scenario 4 as an example.}
    \label{fig:Limitation}
\end{figure}

\section{Route-dependent Attraction-based Stochastic Process Day-to-day Model}\label{sec:Model}

Some existing experiment-based studies attribute the oscillations in route flow to subjects' bounded rational behavior \citep{Iida1992,Selten2007a,Dixit2014,Zhao2016c}.
Although bounded rationality or perceptive errors may partially explain the oscillations, 
this paper attempts to reproduce the oscillations by using stochastic process models based on travelers' independent choice decisions.

Suppose that all travelers are homogeneous \corref{assumption}, and thus the switching rate $p^t_{ij}$ can be viewed as the probability of an individual traveler's switching to route $j$ from route $i$ at time $t$. 
\textcolor{black}{
If a traveler on route $i$ chooses route $j$, it is a 'success' trail with probability $p^t_{ij}$; 
if the traveler chooses other route, it is a 'failure' trial with probability $1-p^t_{ij}$.
Since travelers make decisions independently, the number of the travelers who
switch from route $i$ to route $j$ at time $t$ is the sum of $f_{i}^t$ independent Bernoulli trials with successful rate $p_{ij}^t$. 
}
Therefore, the switching rate $\Delta f_{ij}^t$ in Equation \ref{equ:hit+1} is a stochastic variable that obeys a Binomial distribution and it can be written as follows.
\begin{equation}\label{eq:switching flow}
   \Delta f_{ij}^t \sim \text{Binomial} (f_{i}^t, {p}_{ij}^t).
\end{equation}

\cortext[assumption]{This assumption is not completely true since individuals are heterogenous, no matter in laboratory experiments or in reality. However, it works.}

According to Equation \ref{equ:hit+1}, the flow on route $j$ at time $t+1$, i.e., $f_{j}^{t+1}$, is a sum of Bernoulli-distributed variables from $\Delta f_{1j}^t$ to $\Delta f_{Nj}^t$, and then we have the RDAB-SP model as follows.

\begin{equation}\label{eq:switch flow}
    \begin{cases}
        f_{j}^{t+1} = \sum_{i=1}^N \Delta f_{ij}^t,\\
        \Delta f_{ij}^t \sim \text{Binomial} (f_{i}^t, {p}_{ij}^t),\\
        p^t_{ij}=
        \begin{cases}
        \begin{aligned}
        &P_i \frac{e^{-\theta C_j^t}}{\sum_{k=1}^N e^{-\theta C_k^t}}\ , \ & \ \text{if}\ \ i\neq j\\
        &(1-P_i) + P_i\frac{e^{-\theta C_j^t}}{\sum_{k=1}^N e^{-\theta C_k^t}}\ , \ & \ \text{if}\ \ i=j
        \end{aligned}
        \end{cases}
    \end{cases}
\end{equation}
where $f_{j}^{t+1}$ that is a sum of $N$ independent Binomial-distributed variables follows the Possion-Binominal distribution, and it is determined by and only by the network flow assignment at time $t$, i.e., $\vec f^t$.
Thus, the fact that $\vec{F}^{t+1}$ depends on and only on $\vec{F}^t$ indicates that the evolution of the network flow is a Markov process.
\textcolor{black}{
In the given Markov process, the transition probabilities between any two states are non-zero. 
Thus, it is an irreducible and aperiodic Markov chain with finite state space and a unique equilibrium distribution exists for the Markov process \citep{serfozo2009basics}.
}

The RDAB-SP model in Equation \ref{eq:switch flow} naturally extends the deterministic model in Part I \citep{QI2023103553} of the research by considering individual traveler's independent choice behavior.
On one hand, the RDAB-SP model is closely related to the deterministic model, i.e.,
the switching flow $\Delta f_{ij}^t$ in Equation \ref{equ:hit+1} is the expected value of $\Delta f_{ij}^t$ in Equation \ref{eq:switch flow}.
Therefore, the deterministic model can be seen as the expected or mean evolutionary version of the RDAB-SP model. \textcolor{black}{For the same reason, either the proposed deterministic model or the stochastic model shares the same log likelihood function, i.e., Equation (27) in Part I of the research \citet{QI2023103553}. 
Thus, in the next section of model test, we apply the calibrated parameters of the deterministic model in Table \ref{tab:calibration}, which exhibits the satisfactory generality of the deterministic route-choice framework proposed in Part I of the research \citep{QI2023103553}.}
On the other hand, the Markov property of the RDAB-SP model ensures a non-deterministic evolutionary process, which is obviously different from the deterministic model while much closer to the experimental observations.

\textcolor{black}{
Note that travelers' decisions in reality may not rely on only the latest travel costs. 
It seems to be paradoxical with the `non-aftereffect' property of Markov chain models. 
Indeed, most previous laboratory experiments suggest substantive heterogeneity in the frequency of switches \citet{Qi2019},\citet{Selten2007a}, indicating that some subjects respond to only the latest incentives, while others take more historical experiences into consideration. 
In contrast, this paper proposes a Markov chain model that jointly considers the above two decision-making processes. 
First, if the Markov chain model, which requires less historic information than the classic reinforcement learning models or typical DTD models, could predict the dynamics of human subjects well, we could get a dynamics model as simple as possible. 
Second, because of its minimal demand for historical information, the proposed Markov chain model could start forecasting the aggregated traffic flow at any time without a cold start process, and thus could be utilized in practice.}

\section{Model Test}\label{sec:Test}

This section tests the predictability of the proposed RDAB-SP model by using three indices, namely the switching flow (Equation \ref{eq:switching flow}), the one-step flow evolution (Equation \ref{eq:switch flow}), and the equilibrium flow distribution \citep{cantarella1995dynamic,smith2014long}. 
Besides, the generality of the proposed  RDAB-SP model is examined by the experimental data collected by other researchers.

\subsection{Switching Flow}\label{sec:SwtichingFlow}

As assumed by the proposed model, the switching rate from route $i$ to route $j$ is determined by the current assignment of the network flows. 
Given a specific flow assignment $\vec f^*=(f_1^*, f_2^*, \cdots, f_N^*)$ and if an experimental data record $\vec f^t$ is equal to $\vec f^*$, $\Delta f_{ij}^t$ that is determined by $\vec f^t$ is a sample of the random variable $\Delta f_{ij}^*$ determined by $\vec f^*$.
Therefore, a group of samples can be constructed by calculating $\Delta f_{ij}^t$ at time $t$ where $\vec f^t=\vec f^*$, and the sample size depends on the number of the realizations of $\vec f^*$ in the experiments. 
According to Equation \ref{eq:switching flow}, 
the sample is supposed to follow a Binomial distribution with trial number $f_i^*$ and successful rate $p_{ij}^*$.

To validate the RDAB-SP model, we compare the distribution of the switching flow in the experiments with the distribution derived from the proposed model.
Results are presented in Figure \ref{fig:sfdistributions}. 
The figure of each scenario consists of two panels. 
The left panel is the comparison of the mean and standard deviation between the experimental and theoretical distributions of the switching flow, which are calculated as follows. 
\begin{equation}
    \begin{cases}
        \mu(\Delta f_{ij}) = f_{i} p_{ij}\\
        \sigma(\Delta f_{ij}) = \sqrt{f_{i} p_{ij} (1-p_{ij}})
    \end{cases}
\end{equation}
The right panel shows the detailed comparison of the distribution curves at a representative flow assignment\footnote{The detailed distributions of all cases cannot be presented here due to space limitation}, accompanied by the switching rate and the result of the KS test.

\begin{figure}[!htbp]\ContinuedFloat*
    \centering
    \subfigure[Scenario 1]{
    \includegraphics[width=5in]{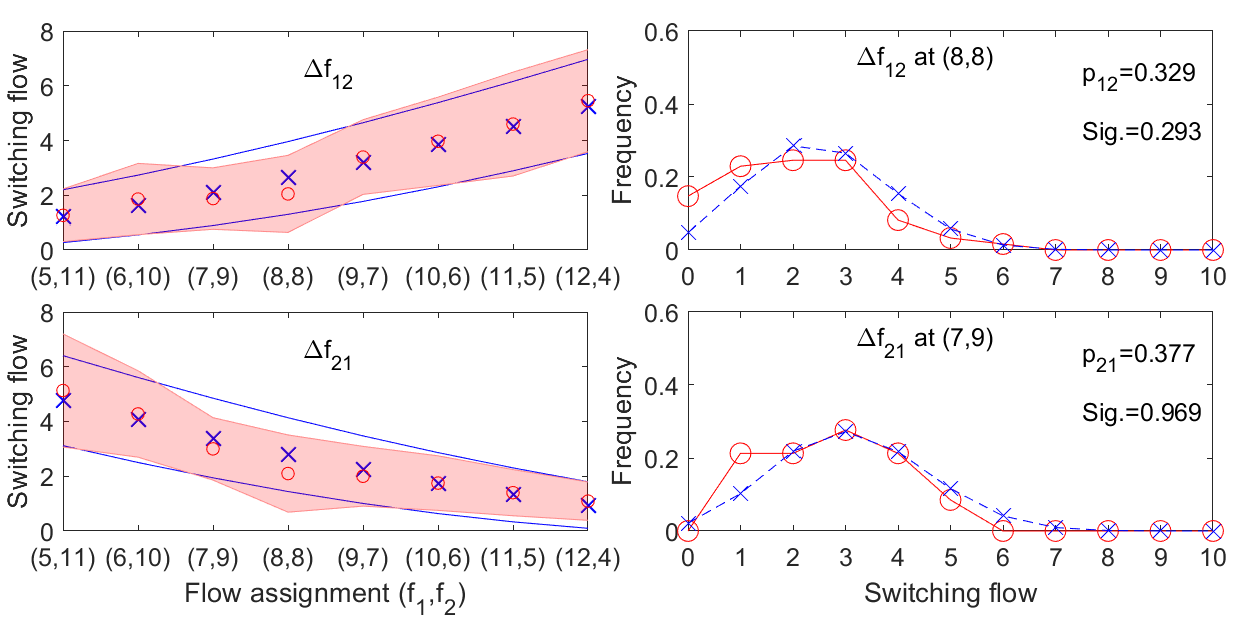}}
    \subfigure[Scenario 2]{
    \includegraphics[width=5in]{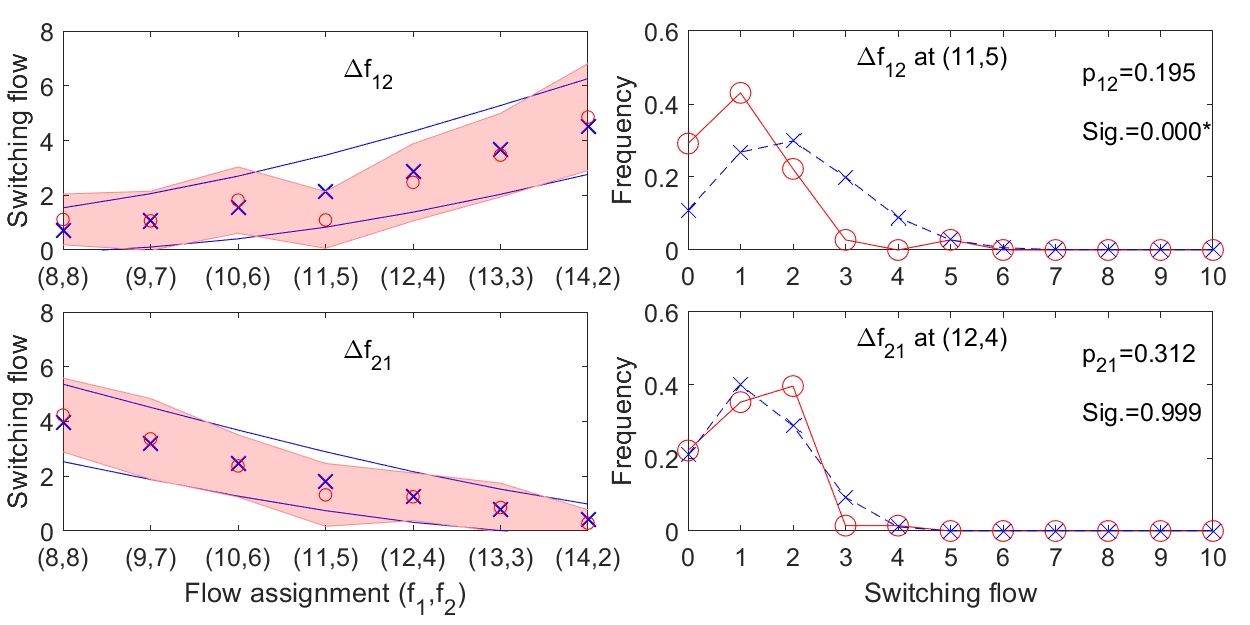}}
    \subfigure[Scenario 3]{
    \includegraphics[width=5in]{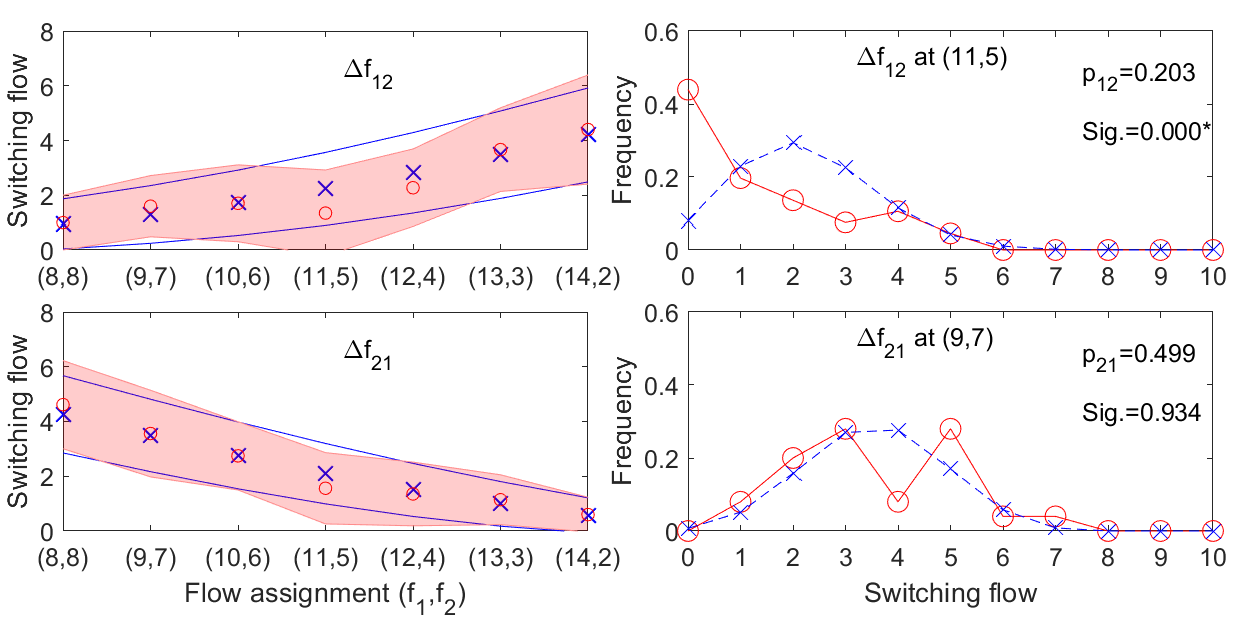}}
    \caption{Comparison between the experimental and theoretical means and standard deviations of switching flow.}
\label{fig:sfdistributions}    
\end{figure}

\begin{figure}[!htbp]\ContinuedFloat
    \centering
    \vspace{-10mm}
    \subfigure[Scenario 4]{
    \includegraphics[width=5in]{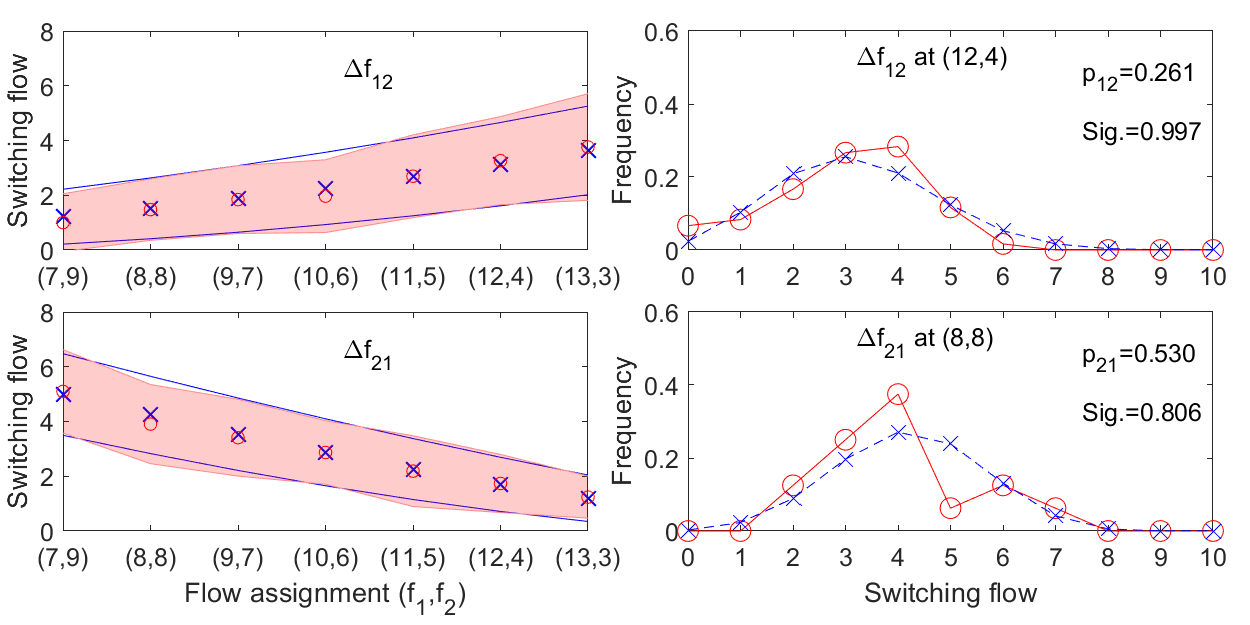}}
    \subfigure[Scenario 5]{
    \includegraphics[width=5in]{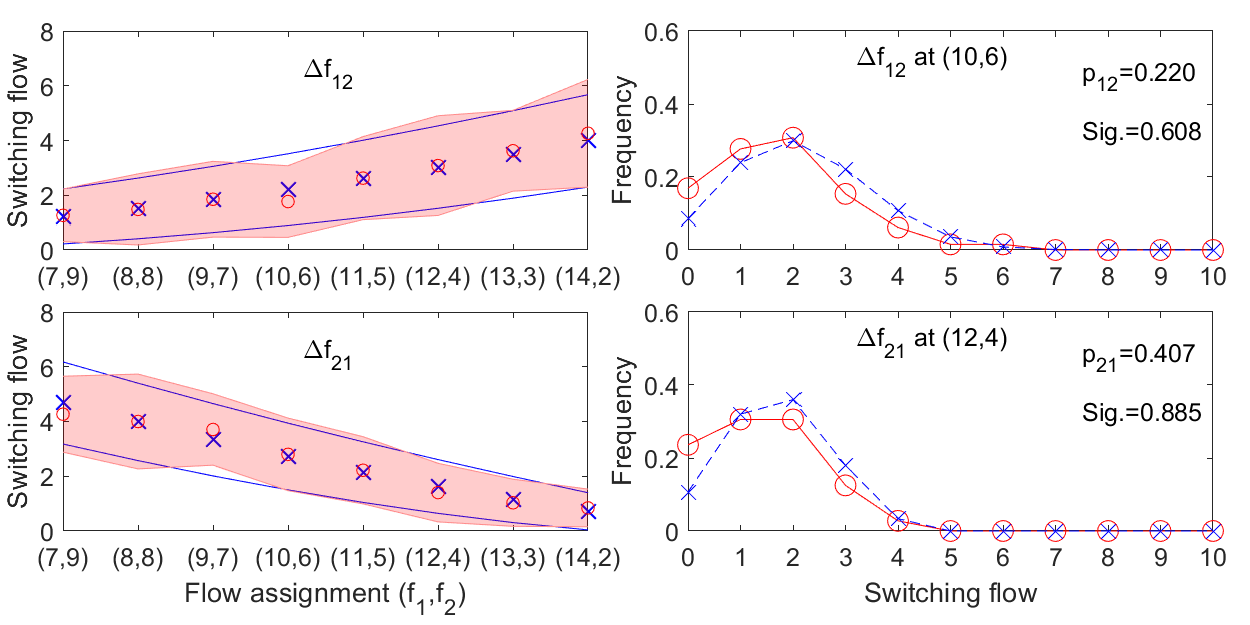}}
    \subfigure[Scenario 6]{
    \includegraphics[width=5in]{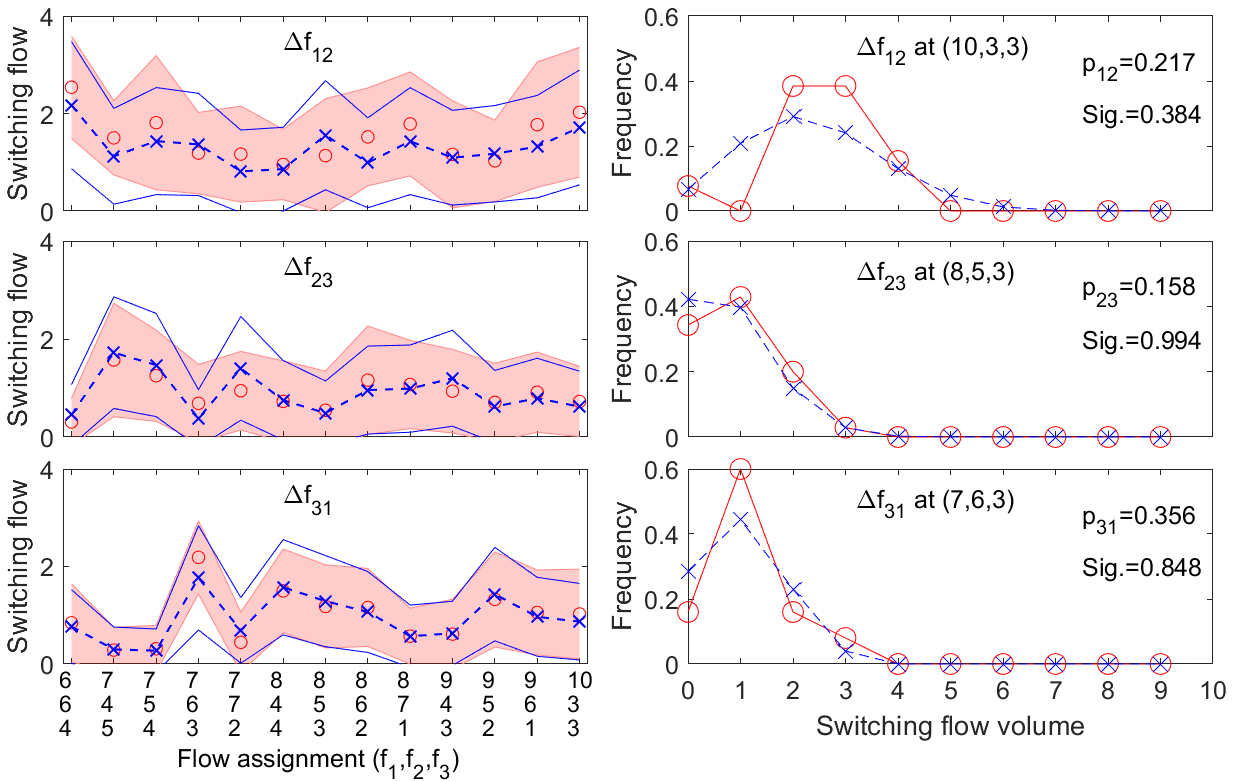}}
    \caption{Continued.}
\label{fig:sfdistributions}    
\end{figure}

\begin{figure}[!htbp]\ContinuedFloat
    \centering
    \subfigure[Scenario 7]{
    \includegraphics[width=5in]{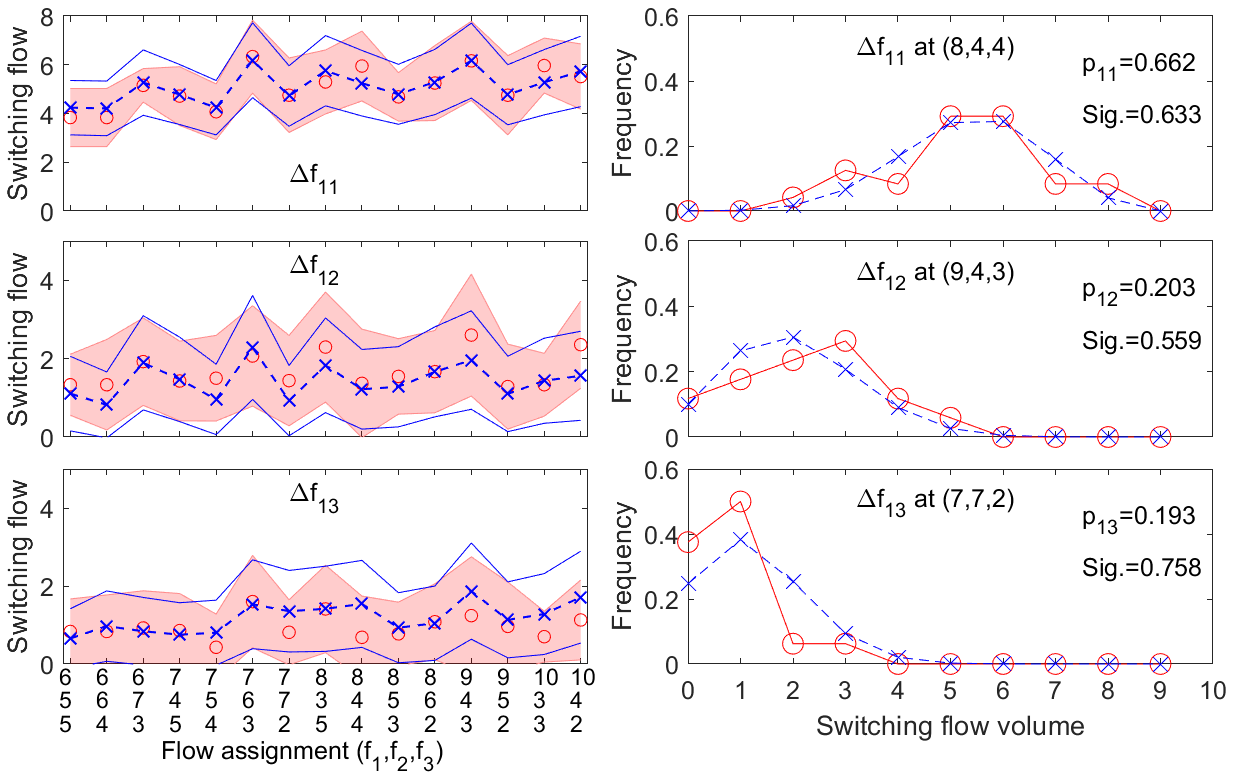}}
    \subfigure[Scenario 8]{
    \includegraphics[width=5in]{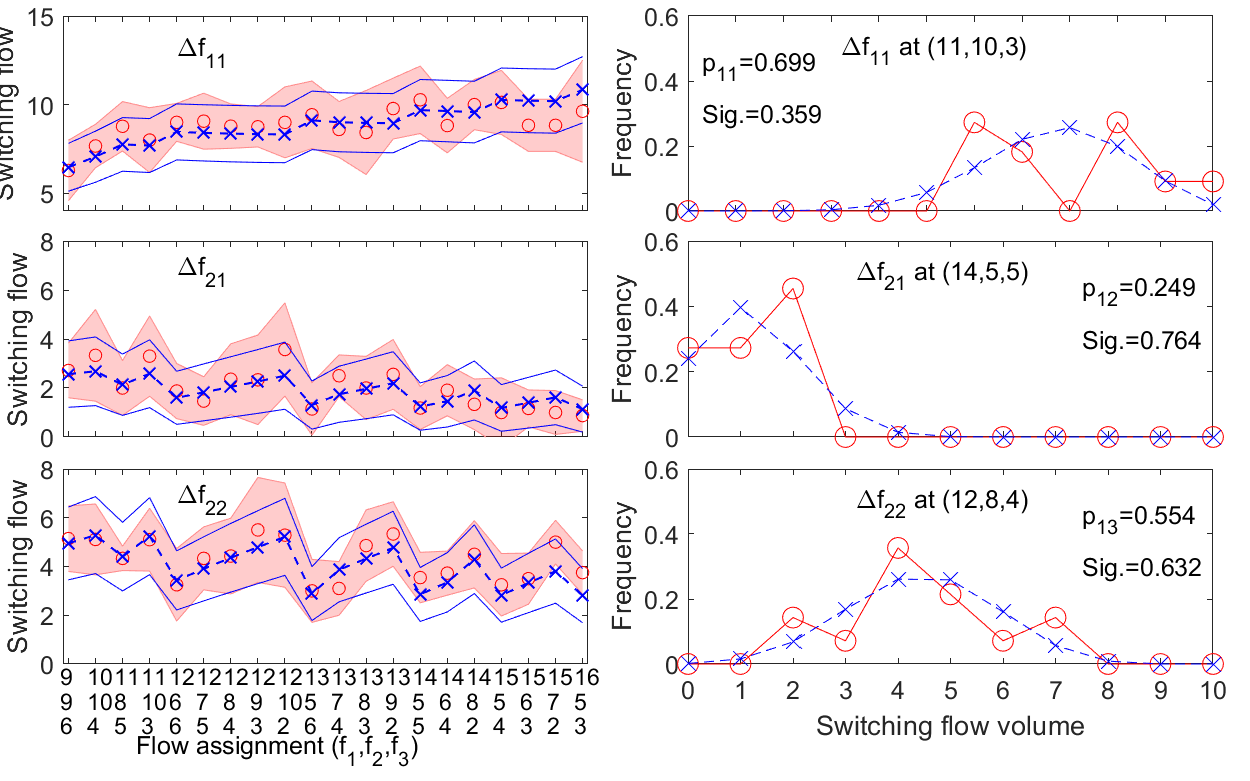}}
    \subfigure{
    \includegraphics[width=5in]{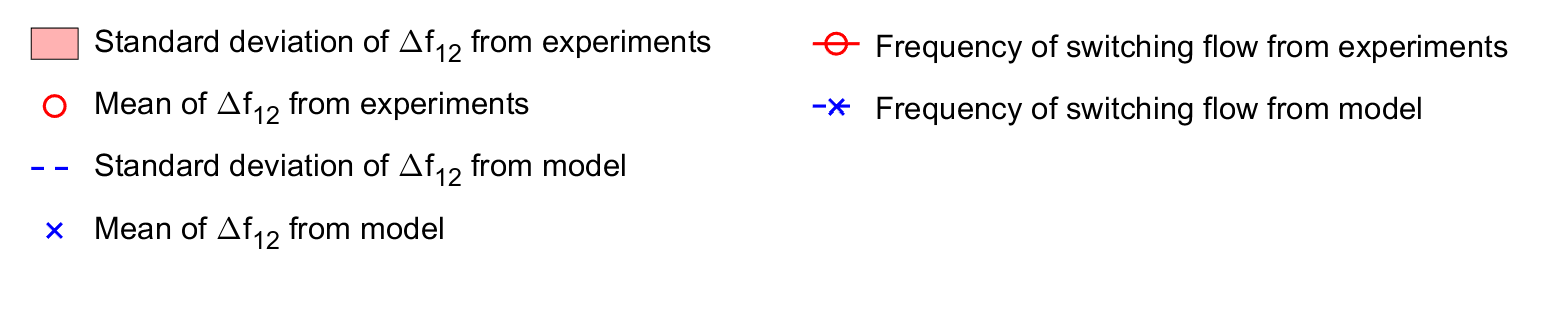}}
    \caption{Continued.}
\end{figure}

It can be seen from Figure \ref{fig:sfdistributions} that the model performs well in describing the switching flow.
Specifically, in the left panels, most of the means and standard deviations resulted from the proposed model well fit the experimental results.
The detailed distributions in some cases where the model doesn't perform very well are selected and presented in the right panels, and it can be seen that the shapes of the distribution curves are still similar to each other, further confirming a good performance of the proposed RDAB-SP model. 
The KS test results also indicate no significant difference, except the user equilibrium points in Scenarios 2 and 3 (actually they are the only exceptions).
We speculate that travelers' behavior patterns might be changed at user equilibrium. 
In-depth investigation is out of the scope and we will carry it out in the future.

\subsection{One-step Flow Evolution}
The stochastic process defined in Equation \ref{eq:switch flow} indicates that $f_j^{t+1}$ follows a Possion-Binominal distribution characterized by $\vec f^*$, given a flow assignment $\vec f^t=\vec f^*$. 
This subsection tests if the experimental data is consistent with such a theoretical prediction.

\begin{figure}[!htbp]
    \centering
    \includegraphics[width=6.6in]{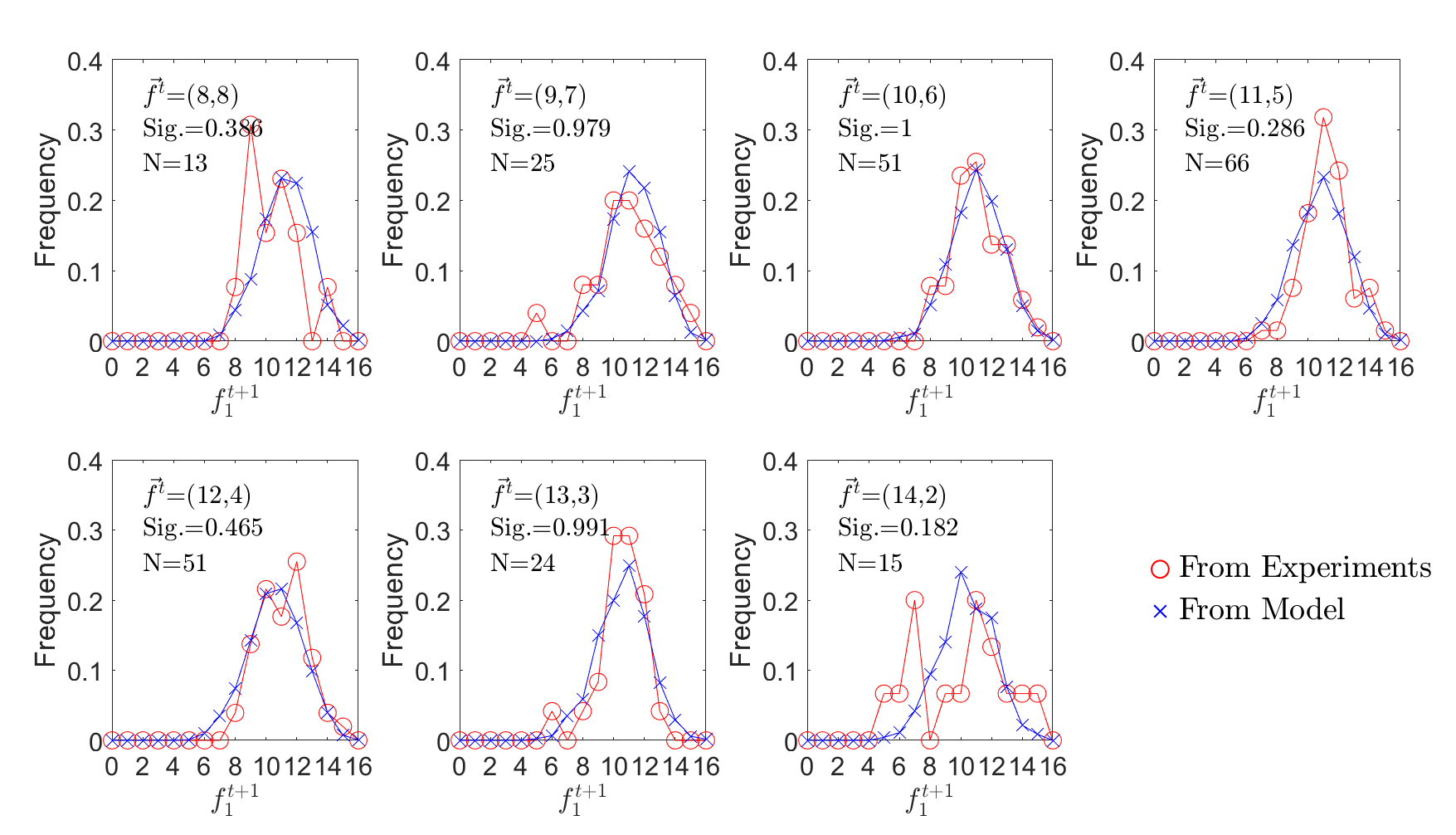}
    \caption{Comparison between the experimental and theoretical one-step flow evolution of the route 1 in Scenario 3. $N$ is the sample sizes and Sig. is the result of the KS test.}
    \label{fig:S3F2F}
\end{figure}

\begin{figure}[!htbp]
    \centering
    \includegraphics[width=6.6in]{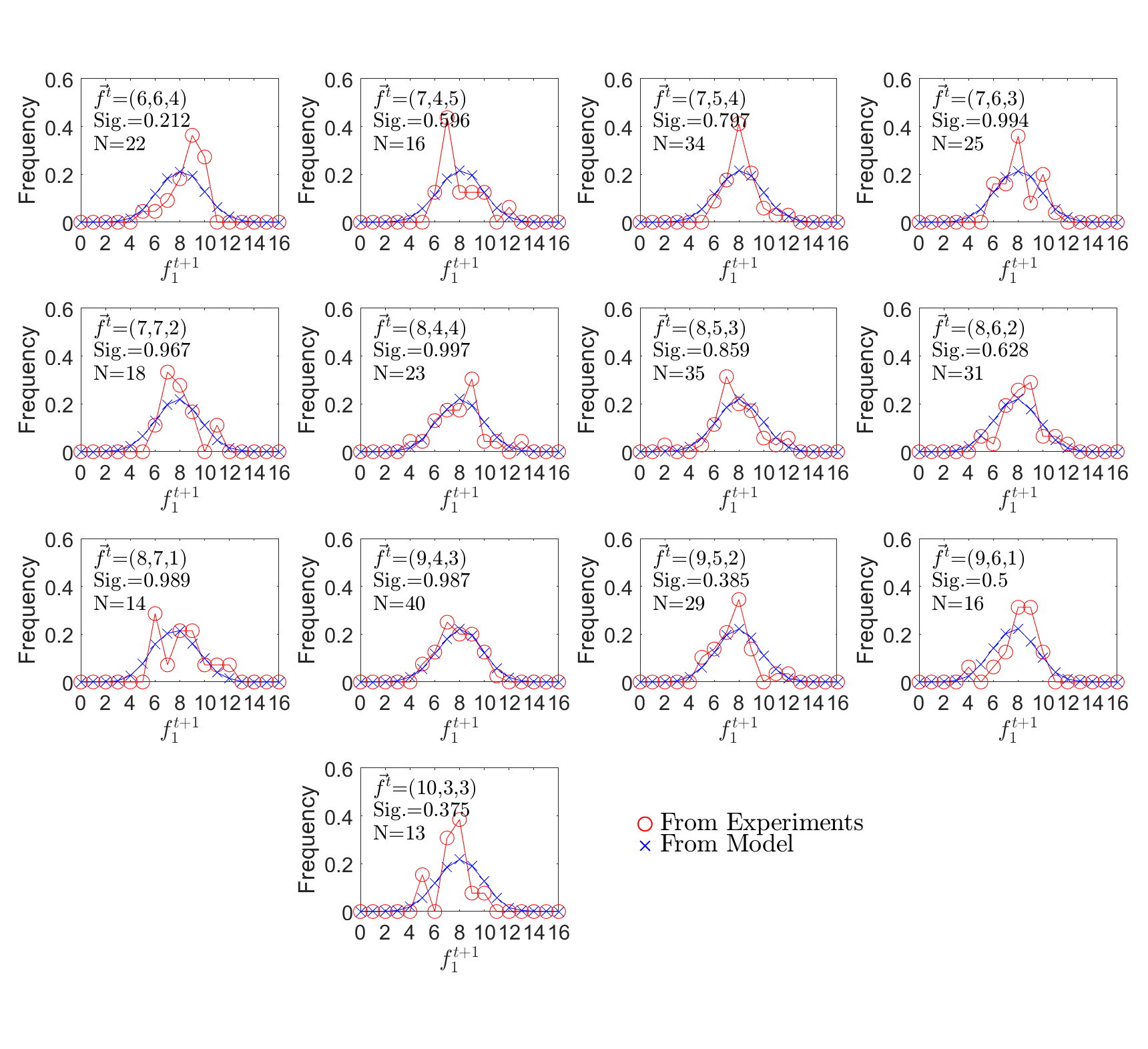}
    \caption{Comparison between the experimental and theoretical one-step flow evolution of the route 1 in Scenario 6. $N$ is the sample sizes and Sig. is the result of the KS test.}
    \label{fig:S6F2F}
\end{figure}

First, the experimental distributions of $f_1^{t+1}$ that corresponds to specific $\vec f^t$ are extracted by using the method similar to that in Section \ref{sec:SwtichingFlow}, and then they are compared with the theoretical distributions predicted by using the proposed model. 
Routes 1 in Scenarios 3 (two-route scenario) and 6 (three-route scenario) are presented as examples in Figures \ref{fig:S3F2F} and \ref{fig:S6F2F}, respectively.

It can be found from Figure \ref{fig:S3F2F} that the shapes of the distribution curves are similar to each other in Scenario 3 and none of the KS test results is significant, turning out that there is no significant differences between the experimental and theoretical distributions.

The deviation between the experimental and theoretical distributions becomes larger in Scenario 6 (Figure \ref{fig:S6F2F}), probably due to the fact that the sample size is much smaller in the three-route scenario\footnote{Compared with the two-route scenario, the sample size is smaller in the three-route scenario, as there are much more flow combinations in the three-route scenarios.}. 
However, the two distributions are still visually similar in particular when the sample size is relatively large, and the KS test also indicates no significant difference.
Therefore, the proposed RDAB-SP model could well explain the experimental data in terms of one-step evolution.

\subsection{Route Flow Distribution}

\begin{figure}[!htbp]
    \centering
    \subfigure[Evolution process]{
    \includegraphics[height=2in]{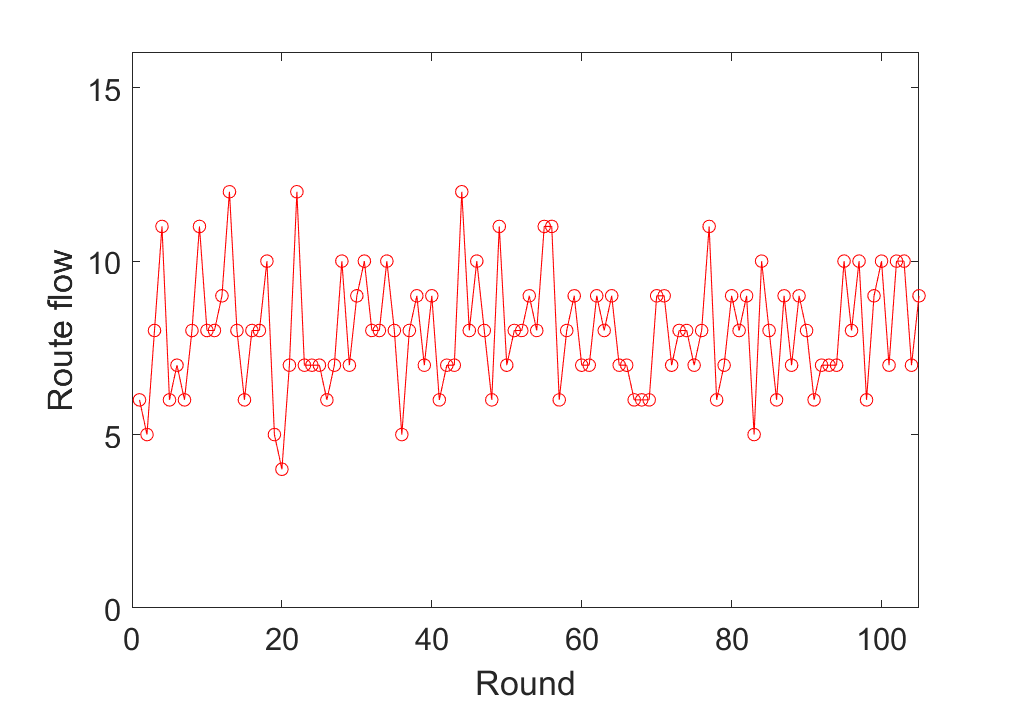}}
    \hspace{5mm}
    \subfigure[Stabilized route flow distribution]{
    \includegraphics[height=2in]{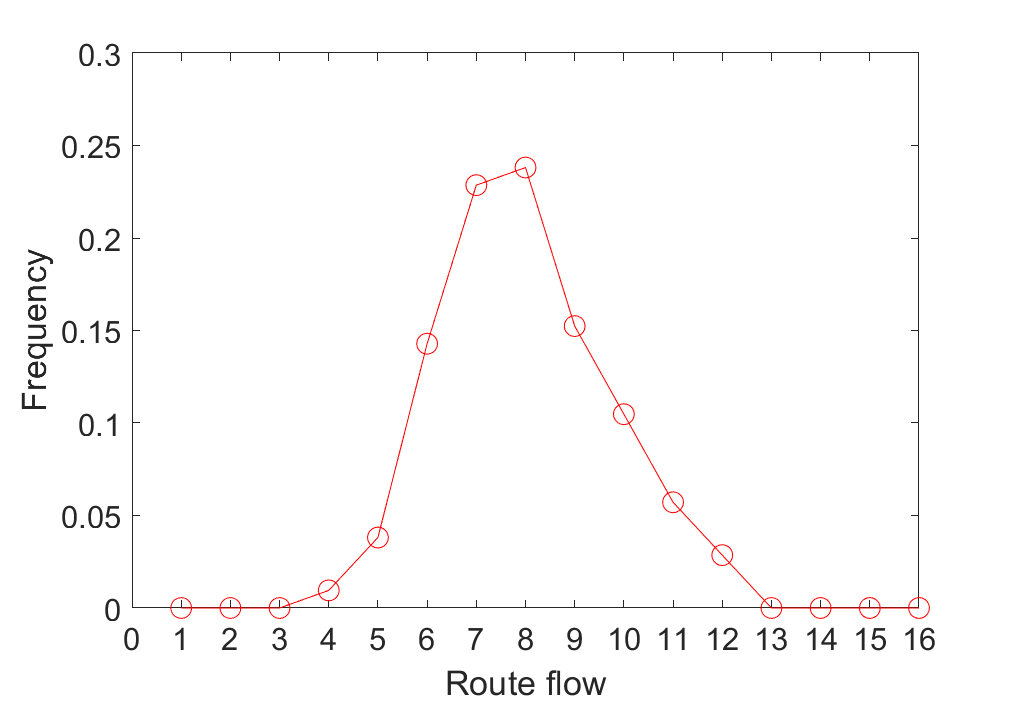}}
    \caption{An example of route flow evolution and distribution (Session 3 in Scenario 2).}
    \label{fig:Exp_Distribution}
\end{figure}

The RDAB-SP model predicts an probability distribution of the equilibrium state in network flows, instead of an specific equilibrium point, turning out that the frequency of the route flow stably stays at a given value in the long run.
To illustrate the existence of the equilibrium distribution, we plot the evolution process and the frequency distribution of the experimental data in Figure \ref{fig:Exp_Distribution}. 
The shape of the distribution curve becomes stable if the evolution process lasts for a long period.

We compare the route flow distributions in the experiments with those derived from the proposed model. 
Different from the switching rates, the theoretical values of the equilibrium route flow distributions are difficult for analytical calculations, since the Markov process is complex to solve. 
Thus, we carry out numerical simulations to obtain the theoretical flow distributions.

\begin{figure}[!htbp]
    \centering
    \subfigure[Scenario 1]{
    \includegraphics[width=3in]{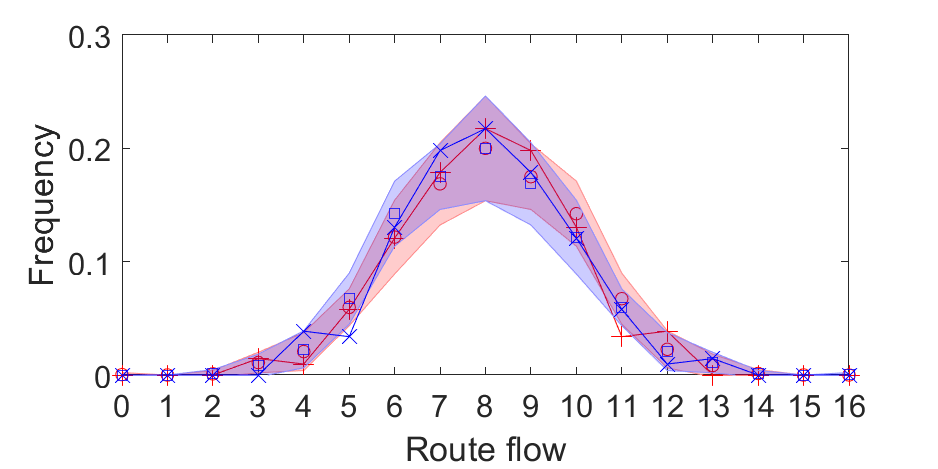}}
    \subfigure[Scenario 2]{
    \includegraphics[width=3in]{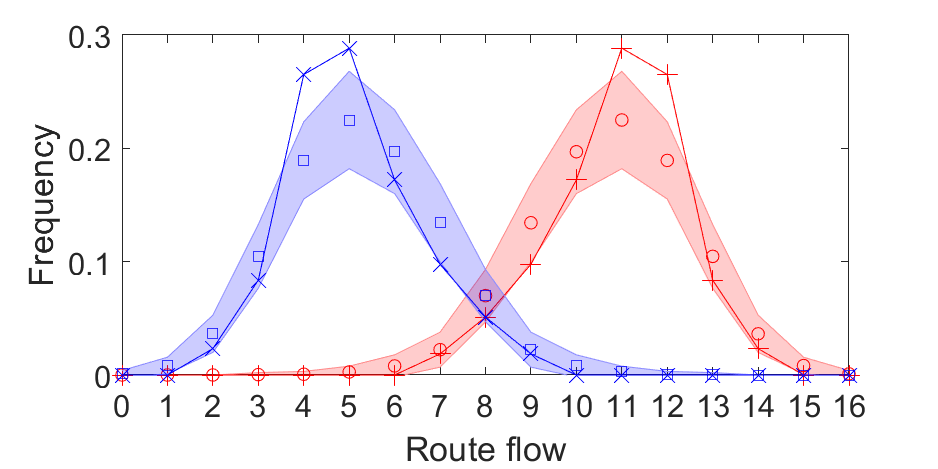}}
    %
    \subfigure[Scenario 3]{
    \includegraphics[width=3in]{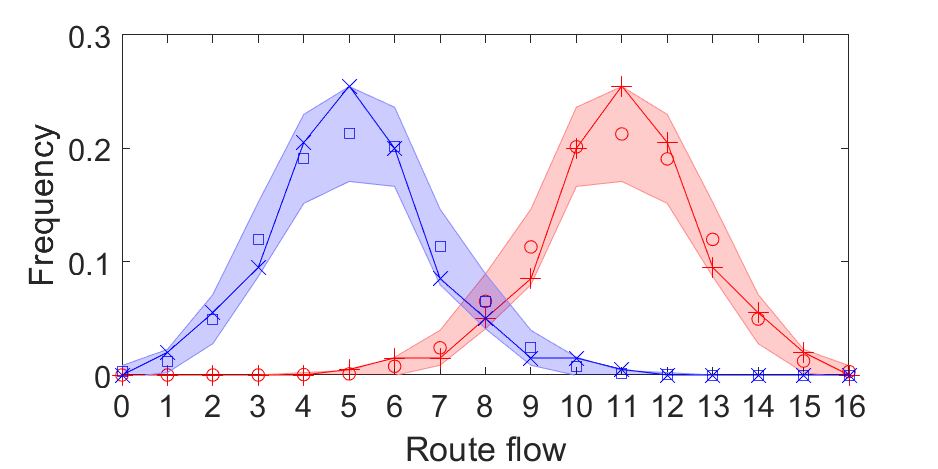}}
    \subfigure[Scenario 4]{
    \includegraphics[width=3in]{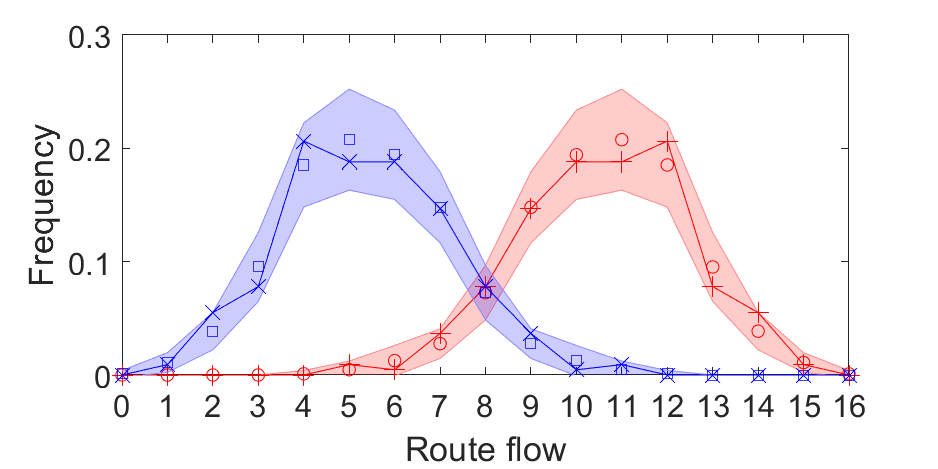}}
    %
    \subfigure[Scenario 5]{
    \includegraphics[width=3in]{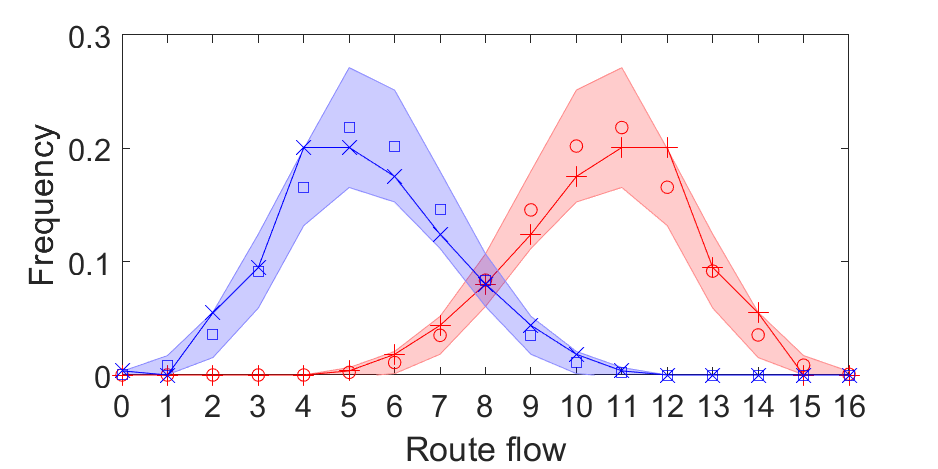}}    
    \subfigure[Scenario 6]{
    \includegraphics[width=3in]{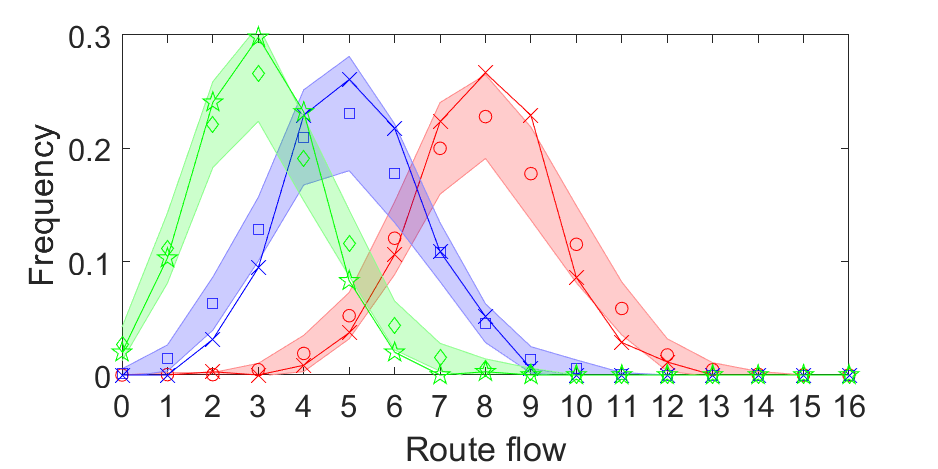}}
    %
    \subfigure[Scenario 7]{
    \includegraphics[width=3in]{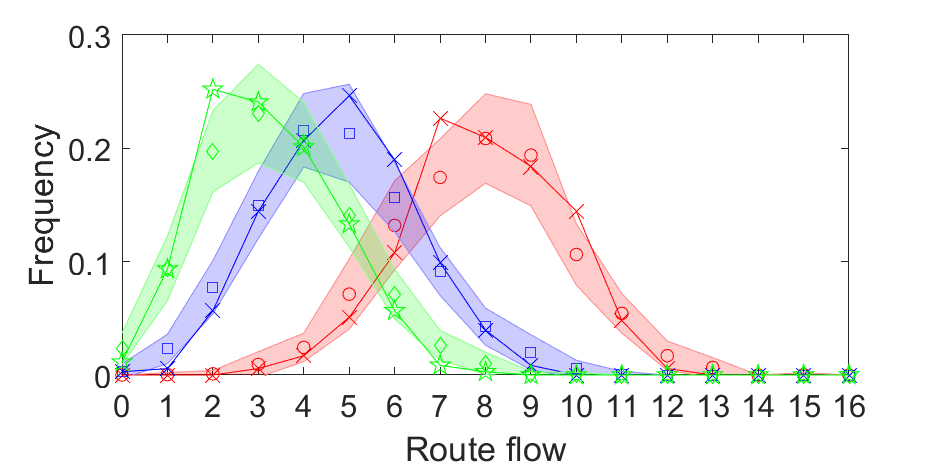}} 
    \subfigure[Scenario 8]{
    \includegraphics[width=3in]{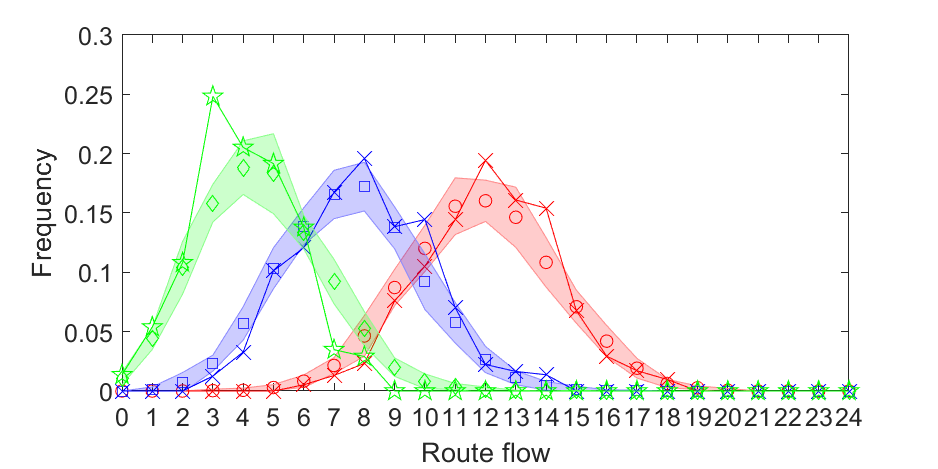}}    
    \subfigure{
    \includegraphics[width=5in]{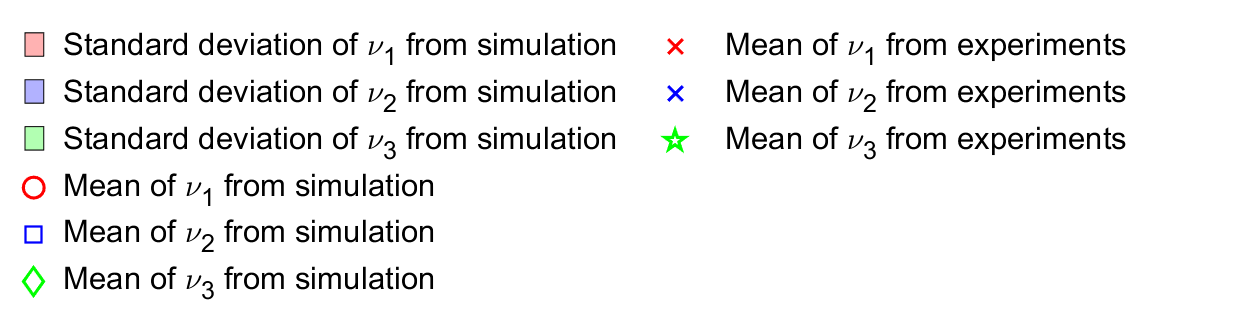}}    
    \caption{The comparison between the experimental and theoretical route flow distributions.}
    \label{fig:Comparison_Distribution}
\end{figure}

To be consistent with the experiment, the configuration of the simulation are set to be the same as that in the corresponding experiment, including the network configuration, the cost functions, the number of travelers, the number of rounds, and the calibrated model parameters.
The simulation is repeated for 30 times for each scenario. 
A group of flow distributional data is thus obtained from every repetition.
Let $\nu_i(x)$ denote the frequency of the case that the flow of route $i$ is $x$, and the mean and standard deviation of $\nu_i$ can be calculated from the 30 groups of flow distributional data.
Results that are presented in Figure \ref{fig:Comparison_Distribution} clearly show that the proposed model well fits the experimental data, indicating that the model is capable to describe the stochastic details of the route flow DTD evolution.

\subsection{Other Data Source}

\begin{figure}[!htbp]
    \centering
    \includegraphics[width=3in]{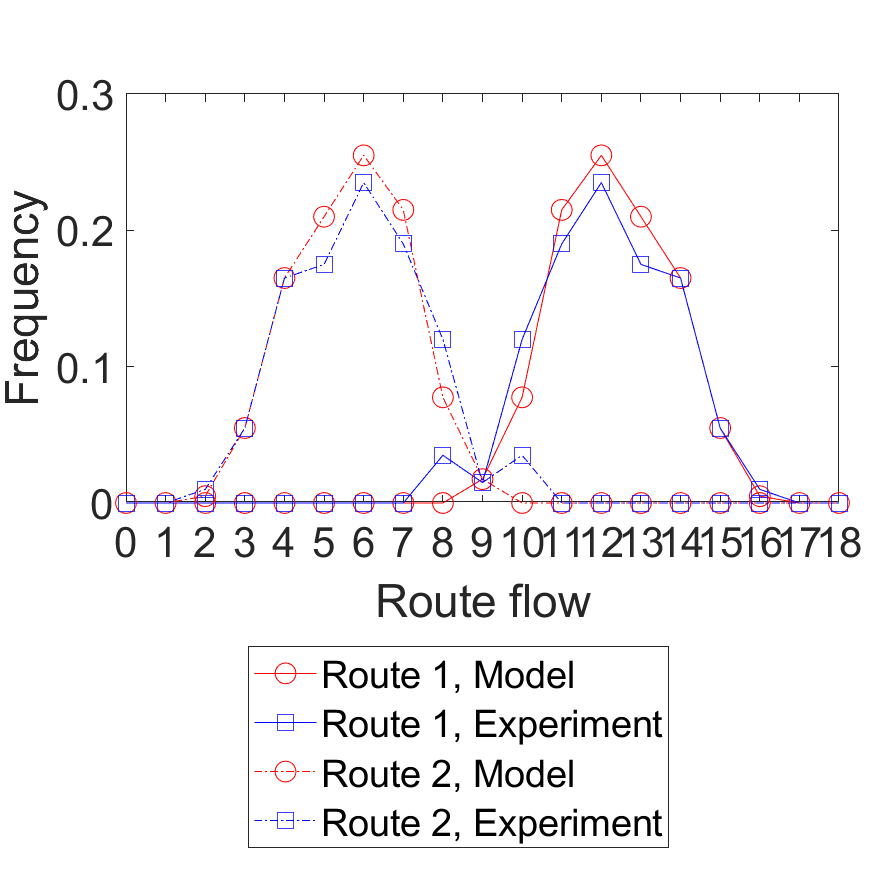}
    \caption{Result comparisons between the model estimation and the experiment data in \cite{Selten2007a}.}
    \label{fig:Selten}
\end{figure}

To examine the model generalization, we validate the RDAB-SP model using experimental data in the existing paper.
In the influential pioneer study \citep{Selten2007a}, a group of subjects (18 for each session) were invited to make a decision from two alternative routes with different cost functions, and the experiments lasted for 200 rounds. 
The route flow distribution is extracted from the 200 experimental data points from Figure 2 in \cite{Selten2007a}.
We calibrate the parameters of the proposed model by using the flow distribution data. 
The following performance indicator is used in the calibration. 
\begin{equation}
    \min \sum_{k=1}^F \sum_{j=1}^N \frac{|\tilde r_{ki} - r_{ki}|}{r_{ki} + \xi }
\end{equation}
where $r_{ki}$ and $\tilde r_{ki}$ are the frequency of the case that subject $k$ selects route $i$ in all 200 rounds derived from the proposed model and the experiment, respectively;
$\xi$ is a small real value to avoid dividing zero. 
This indicator aims to minimize the relative deviation between the experimental and model-estimated flow distributions.
The calibrated parameters are $\theta=0.349$, $\eta_1=0.752$, and $\eta_2=0.683$.
Figure \ref{fig:Selten} compares the results, and it shows that the proposed models achieve high accuracy of prediction.



\textcolor{black}{
In summary, the RDAB-SP model is shown to be able to well explain the stochastic oscillations exhibited by the experimental data, and it overcomes the major limitations of the deterministic model without introducing more assumptions.}

\section{Approximate Model}\label{sec:Approximate}

It has been shown that the RDAB-SP model successfully explains the experimental observations in a stochastic perspective. 
However, sampling a large number of Binomial-distributed random variables results in high computational burden, which hinder the model from practical applications in transportation planning and/or policy design. 
To overcome this issue, this section provides an approximate model to simplify the simulation process with acceptable loss of accuracy.



Recalling Equation \ref{eq:switch flow}, the switching flow $\Delta f_{ij}^t$ follows a Binomial distribution with $n=f_{i}^t$ and $p={p}_{ij}^t$.
Considering the fact that the number of route users is usually quite large, the switching flow $\Delta f_{ij}^t$ could be approximated by a Gaussian distribution\footnote{
According to the continuity correction theory \citep{anscombe1948the}, if $np$ and $np(1-p)$ in a Binomial distribution $B(n,p)$ are large than 5, the Binomial distribution can be fairly well approximated by a Gaussian distribution $N(np, np(1-p))$. 
} as follows.
\begin{equation}\label{eq:switchflowapp}
   \Delta f_{ij}^t \sim \textbf{N}(f_{i}^t{p}_{ij}^t\ ,\  f_{i}^t{p}_{ij}^t(1-{p}_{ij}^t)).
\end{equation}

Since the travelers on different routes choose routes independently, $f_{j}^t$ is the sum of $N$ independent Gaussian-distributed variables, i.e.,
\begin{equation}\label{eq:flowapp}
   f_{j}^{t+1} = \sum_{i=1}^N \Delta f_{ij}^t \sim \textbf{N}\left(\sum_{i=1}^N f_{i}^t{p}_{ij}^t\ ,\  \sum_{i=1}^N f_{i}^t{p}_{ij}^t(1-{p}_{ij}^t)\right).
\end{equation}

Equation \ref{eq:flowapp} can be rewritten as the sum of a fixed term and a random term through the transformation to the standard Gaussian distribution, i.e., {\it the approximate model} of the RDAB-SP model:
\begin{equation}\label{eq:flowapp2}
   f_{j}^{t+1} = \sum_{i=1}^N f_{i}^t{p}_{ij}^t + \sqrt{ \sum_{i=1}^N f_{i}^t{p}_{ij}^t(1-{p}_{ij}^t)} \epsilon^t_j
\end{equation}
where $\epsilon^t_j$ is a random variable following the standard Gaussian distribution $N(0,1)$.
Equation \ref{eq:flowapp2} is the approximate model of the RDAB-SP model in Equation \ref{eq:switch flow}.
It is interesting to see that it is the deterministic model (Equation \ref{equ:hit+1}) plus a random term following a Gaussian distribution.

\subsection{Stability Analysis}\label{sec:Stability}

The accuracy of the approximation is determined by the accuracy of the Gaussian approximation of a Binomial distribution.
Let $\bar f_j^{t+1}$ be the deterministic term of Equation \ref{eq:flowapp2}, i.e.,
\begin{equation}\label{eq:flowapp2deter}
    \bar f_{j}^{t+1} = \sum_{i=1}^N f_{i}^t{p}_{ij}^t, 
\end{equation}
The approximation might be failed if Equation \ref{eq:flowapp2deter} has no stable equilibrium \citep{Davis1993Large}. 
\textcolor{black}{
Therefore, we carry out a stability analysis in the following subsection. }


\textcolor{black}{Prior to a formal proof}, we first introduce \textbf {Lyapunov Theorem of a Discrete Dynamic System} as follows. 
Consider a discrete-time and autonomous dynamic system 
\begin{equation}
    X(t+1)=H(x(t))
\end{equation}
where $H(0)=0$ for all $t$. 
Suppose that there exists a scalar function $V(x)$ such that for $V(0)=0$ for all $t$ and

\begin{itemize}
\vspace{-3 mm}
    \setlength{\itemsep}{0pt}
    \setlength{\parsep}{0pt}
    \setlength{\parskip}{0pt}
    
\item $V(x)>0 \;\text{when}\; x\neq 0 $;
\item $\Delta V(x)<0 \;\text{when}\; x\neq 0$, where $\Delta V(x)$ is the rate of increase of $V$ along the motion starting at $x$;
\item $V(x)\; \text{is continuous in} \;x$;
\end{itemize}

Then, the equilibrium state $x_{\eta}=0$ is uniformly asymptotically stable  and $V(x)$ is a Lyapunov function of the system \citep{Kalman2003Control}.

We employ the same scalar function that was defined in Equation (12) in Part I of the research \citep{QI2023103553}, i.e.,
\begin{equation}
    V(\vec f)=z(\vec f)-\min z(\vec f)
\end{equation}
where 
\begin{equation}
z(\vec f) = \sum_{i=1}^N  \theta\int_0^{f_i} {C_i} \text{d}x
         + \sum_{i=1}^N  f_i  \text{ln}(P_i f_i)
\end{equation}

\noindent According to the definitions, $V(\vec f)$ meets the first and third conditions in Lyapunov Theorem for a Discrete Dynamic System.
The second condition is proved as follows.
\begin{equation}
\label{eq:stability part I}
\begin{split}
\Delta V(\vec f)
&=
V(\vec f^{t+1})-V(\vec f^t)\\
&=
\sum_{i=1}^N 
\{
\theta (\int_{0}^{f^{t+1}}{C_i^{t+1}}\text{d}x-\int_{0}^{f^{t}}{C_i^{t}}\text{d}x)
+ [f_i^{t+1}  \text{ln}(P_i^{t+1} f_i^{t+1}) - f_i^{t}  \text{ln}(P_i^{t} f_i^{t})]
\}\\
&=
\sum_{i=1}^N 
\{
\theta (\int_{0}^{f^{t+1}}{C_i^{t+1}}\text{d}x+\int_{f^{t}}^{0}{C_i^{t}}\text{d}x)
+ (f_i^{t+1}  \text{ln}(P_i^{t+1} f_i^{t+1}) - f_i^{t}  \text{ln}(P_i^{t} f_i^{t})]
\}\\
&=
\sum_{i=1}^N 
\{
\theta \int_{f^{t}}^{f^{t+1}}{C_i^{t+1}}\text{d}x
+ [f_i^{t+1}  \text{ln}(P_i^{t+1} f_i^{t+1}) - f_i^{t}  \text{ln}(P_i^{t} f_i^{t})]
\}
\end{split}
\end{equation}

By using the Mean Value Theorem for definite integrals and the Lagrange Mean Value Theorem, Equation \ref{eq:stability part I} can be rewritten as follows.
\begin{equation}
\label{eq:stability part II}
V(f^{t+1})-V(f^t) = \sum_{i=1}^N [\theta C_i(\xi_i) + \text{ln}(P_i(\xi_i) \xi_i)+1](f_i^{t+1} - f_i^{t}),
\end{equation}
where $\xi$ is between $f_i^t$ and $f_i^{t+1}$.

Recalling Equation (25) in Part I of the research, it states as follows \citep{QI2023103553}.
\begin{equation}
\label{eq:an old story}
\begin{split}
    G(\vec f^t)
    &=\sum_{i=1}^N[\theta C_i^t + \text{ln}(P_i^t f_i^t)+1]
    \left[ -P_i^t f_i^t + e^{-\theta C_i^t} \frac{\sum_{k=1}^N P_k^t f_k^t}{\sum_{k=1}^N e^{-\theta C_k^t}} \right]\\
    &= \sum_{i=1}^N (\theta C_i^t(f_i^t) + \text{ln}(P_i^t(f_i^t) f_i^t)+1)(f_i^{t+1} - f_i^{t})\\
    &<0.
\end{split}
\end{equation}
Since $G(\vec f^t)$ in Equation \ref{eq:an old story} is continuous at $\vec f^t$, there exists a neighborhood $\mu$ of $\vec f^t$ such that $G(x)<0$ for every $x$ in $\mu$, according to the sign-preserving property. 
Combining Equation \ref{eq:stability part II}, it is known that $V(f^{t+1})-V(f^t)<0$ if $\vec \xi=(\xi_1, \xi_2, \dots, \xi_N)$ is in $\mu$, making the second condition in Lyapunov Theorem for a Discrete Dynamic System to be true. 
\textcolor{black}{
Since $\xi_i$ is between $f^t$ and $f^{t+1}$, $V(f^{t+1})-V(f^t)$ is negative as
long as the difference between $\vec f^t$ and $\vec f_i^{t+1}$ is sufficiently small, followed by the fact that the fixed point of Equation \ref{eq:flowapp2deter} is uniformly and asymptotically stable.
}

Note that it is a sufficient but not necessary condition of stability. 
It requires the route flow not to change dramatically between two successive time steps, which can be guaranteed by using a small learning rate or high choice inertia when choosing the model parameters.

\subsection{Model Comparison}
\begin{figure}[!htbp]
    \centering
    \includegraphics[width=6.5in]{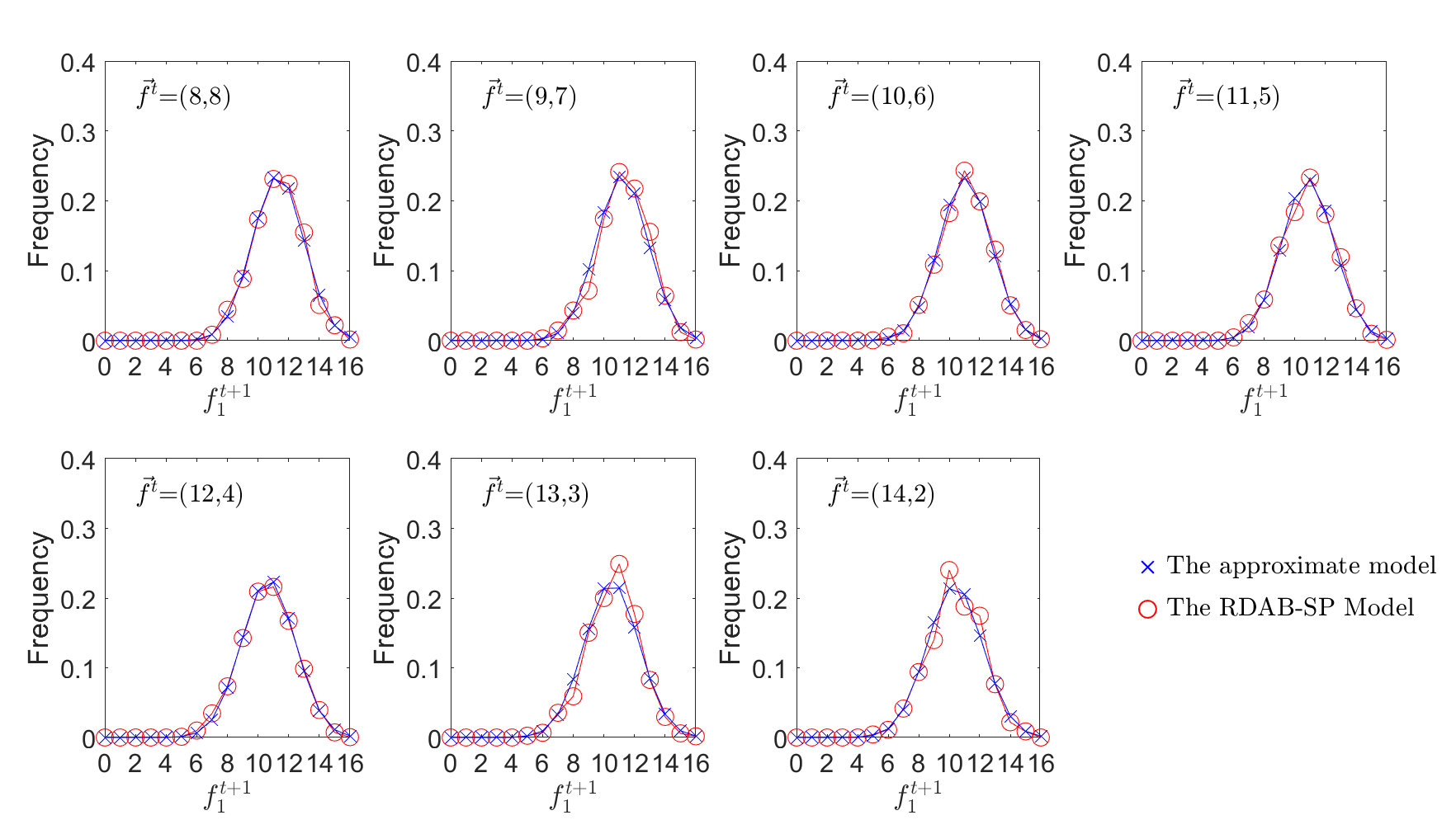}
    \caption{Comparison between the one-step evolution of the route 1 in Scenario 3 generated by the approximate and RDAB-SP models.}
    \label{fig:APPS3F2F}
\end{figure}

\textcolor{black}{
This subsection compares the approximate model with the proposed RDAB-SP model in terms of estimation accuracy on the one-step evolution.}
To save space, we take the route 1 of Scenario 3 as an example and demonstrate the performance of the approximate model in Figure \ref{fig:APPS3F2F}.
The one-step evolution is used as the performance indicator. 
For the one-step evolution, only small deviation occurs in the extreme cases such as $\vec f^t=(13,3)$ and $\vec f^t=(14,2)$, probably resulted from the small number of the users of route 2. 
%
For the equilibrium distribution, the approximate model is also consistent with the RDAB-SP model.
Therefore, it is concluded that the approximate model is a good approximation of the proposed RDAB-SP model.

\section{Numerical examples and policy implications}

This section consist of the following four parts.
The first part (Section \ref{sec:fdist}) examines the influence of parameter changes on equilibrium distributions and the consistency between the RDAB-SP model and the approximate model under different settings. 
The second and third parts (Sections \ref{sec:evol} and \ref{sec:eff}) demonstrate the model stability in the evolution process and the advantage of the approximate model in computing time, respectively.
Policy implications are discussed in the last part (Section \ref{sec:pol}).

Illustrative numerical examples with a three-route road network\footnote{A complex network is not employed because it will not change the results that we intend to examine.} are carried out. 
The travel cost function of each route is quadratic, calculated by using the function $t_i=\alpha_i(1+(f_i/c_i)^2)$ where $c_i$ is the capacity of route $i$. 
The network configuration is set as $\alpha_1=\alpha_2=\alpha_3=3$, $c_0=30$, $c_1=20$ and $c_2=10$.


\subsection{\textcolor{black}{Flow Distribution}}\label{sec:fdist}

\begin{figure}[!htbp]
    \centering
    \includegraphics[width=5in]{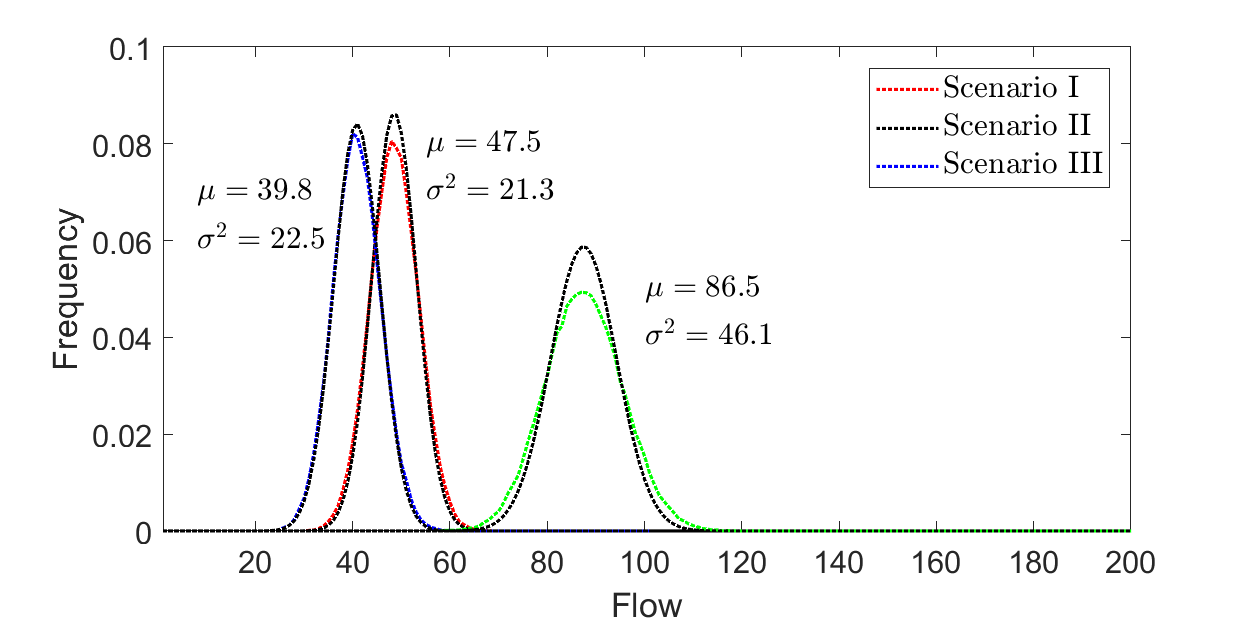}
    \caption{\textcolor{black}{Results of the numerical examples, where the route 1 is focused for simplicity.}}
    \label{fig:Num}
\end{figure}

\textcolor{black}{
Three different simulation scenarios are constructed and employed (Table \ref{tab:numericalexamples}). 
The resulting flow distributions of the route 1 are shown in Figure \ref{fig:Num}, including distribution simulated by using the RDAB-SP model and the corresponding results from the approximate model.}

\textcolor{black}{
The following observations can be obtained from the results in Figure \ref{fig:Num}.
First, the results of the approximate model is very similar to the exact results derived from the RDAB-SP model. 
Second, the route 1 in Scenario I has higher mean flow than that in Scenario II when other parameters are fixed, indicating that the larger route-dependent attraction coefficient is, the more flow volume is attracted.
It is consistent with the results shown in the deterministic model. 
Third, route flow variance grows nearly proportionally with the increase of the demand (by comparing Scenario II with Scenario III) as predicted by Equation \ref{eq:flowapp}.}

\begin{table}[!htbp]\footnotesize\center\textcolor{black}{
\setlength{\tabcolsep}{10mm}{
\caption{\textcolor{black}{Parameters in the numerical examples.}}\label{tab:numericalexamples}
\begin{tabular}{cccccc}
\toprule
{\footnotesize Scenario} & {\footnotesize Demand} & {\footnotesize $\theta$} & {\footnotesize $\eta_1$} & {\footnotesize $\eta_2$} & {\footnotesize $\eta_3$} \\
\midrule
 I & 100 & 0.01 & 0.5 & 0.2 & 0.1 \vspace{2mm}\\
 II & 100 & 0.01 & 0.3 & 0.2 & 0.1 \vspace{2mm}\\
 III & 200 & 0.01 & 0.3 & 0.2 & 0.1 \\
\bottomrule
\end{tabular}}}
\end{table}

\subsection{Evolution Process and Stability}\label{sec:evol}
In this part, the route-dependent attraction coefficients are set as $\eta_0=0.3$, $\eta_1=0.2$ and $\eta_1=0.1$, which are proportional to the capacity;
The total demand $D=100$; The dispersion parameter $\theta$ is to be adjusted to generate different results.

Initially, assume that the numbers of the travelers who select routes 1, 2 and 3 are 0, 50 and 50, respectively.
Then, we run the RDAB-SP model and the approximate model for the next 100 days (rounds). 
For the RDAB-SP model, we simulate the scenario for 30 times, the mean value and 95\% confidential level of each day are calculated based on the 30 data points.
For the approximate model, we estimate the mean of $f_i^{t+1}$ by using Equation \ref{eq:flowapp2deter} and calculate the 95\% confidential level of $f_i^{t+1}$ by assuming that $f_i^{t+1}$ follows a Gaussian distribution with variance $\sum_{i=1}^N f_{i}^t{p}_{ij}^t(1-{p}_{ij}^t)$.
We carry out three groups of the numerical experiments with $\theta=0.01$, $\theta=0.1$ and $\theta=1$, respectively.

\begin{figure}
    \centering
    \subfigure[$\theta=0.01$]{
    \includegraphics[width=5.5in]{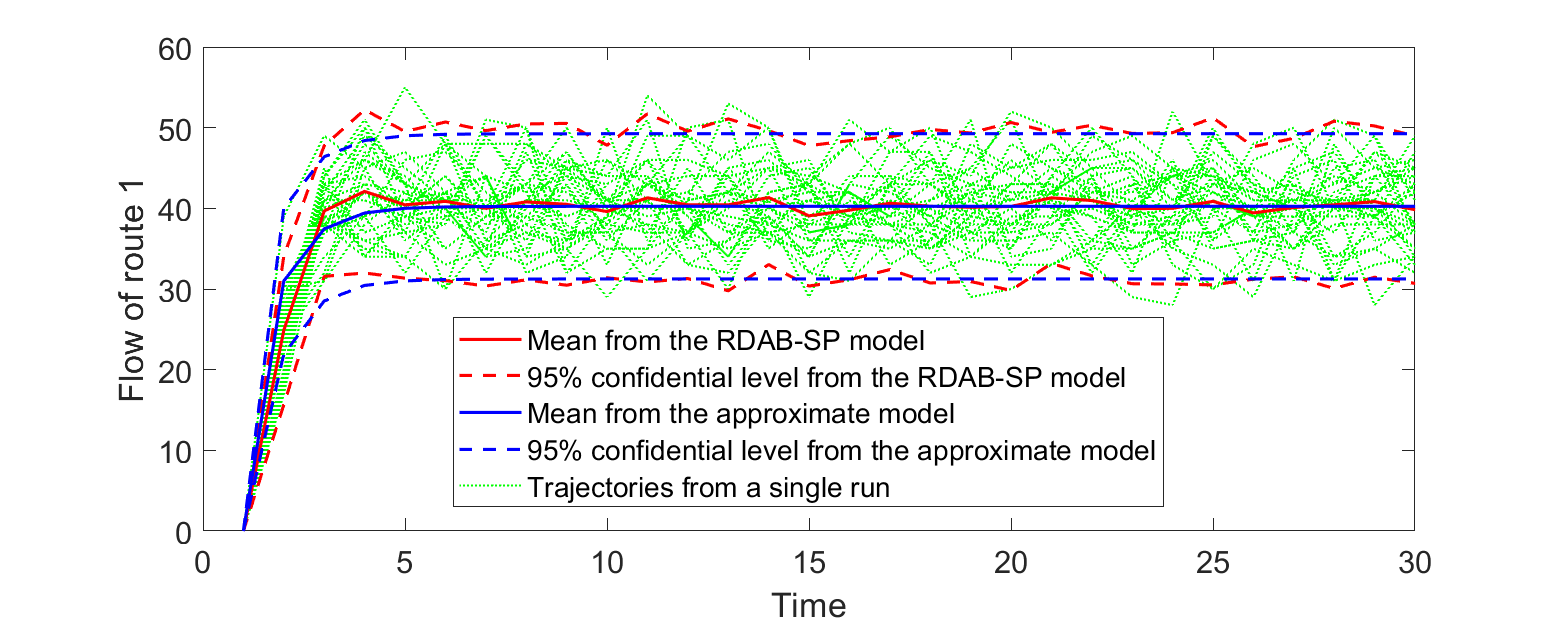}}    
    \subfigure[$\theta=0.1$]{
    \includegraphics[width=5.5in]{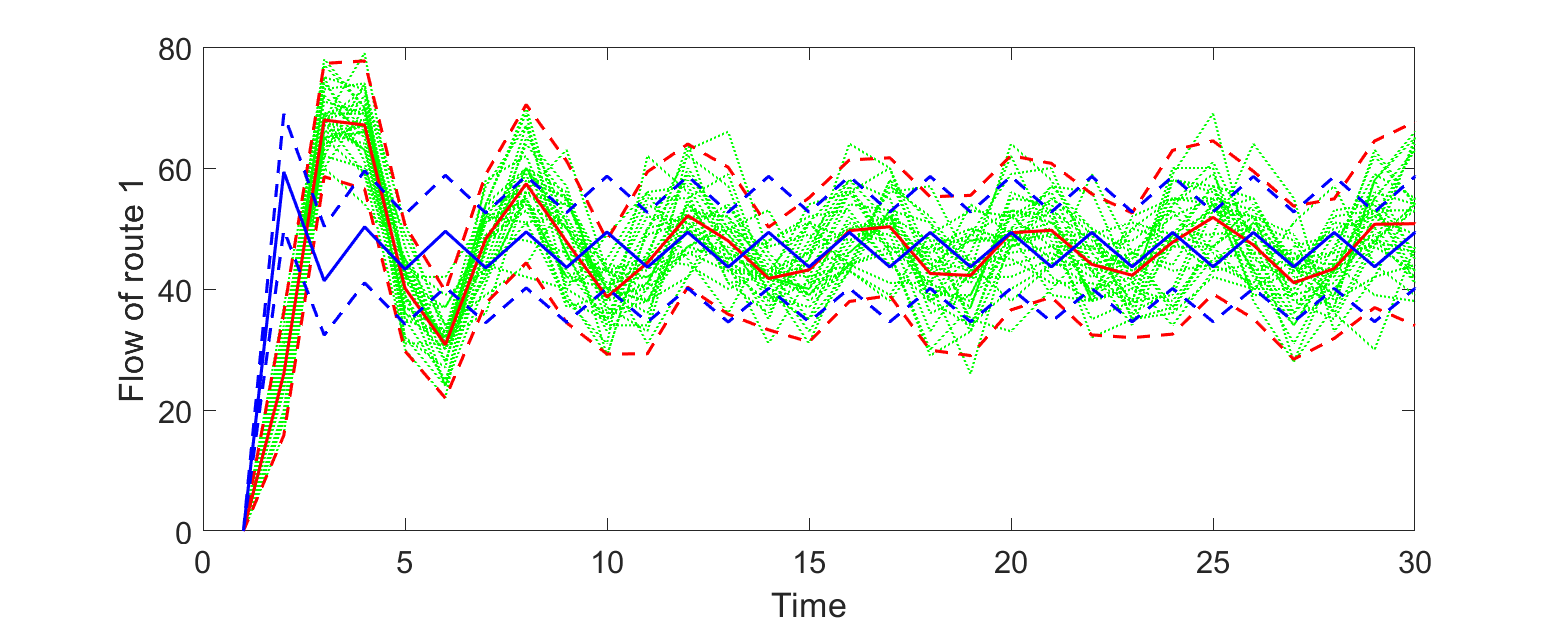}}
    %
    \subfigure[$\theta=1$]{
    \includegraphics[width=5.5in]{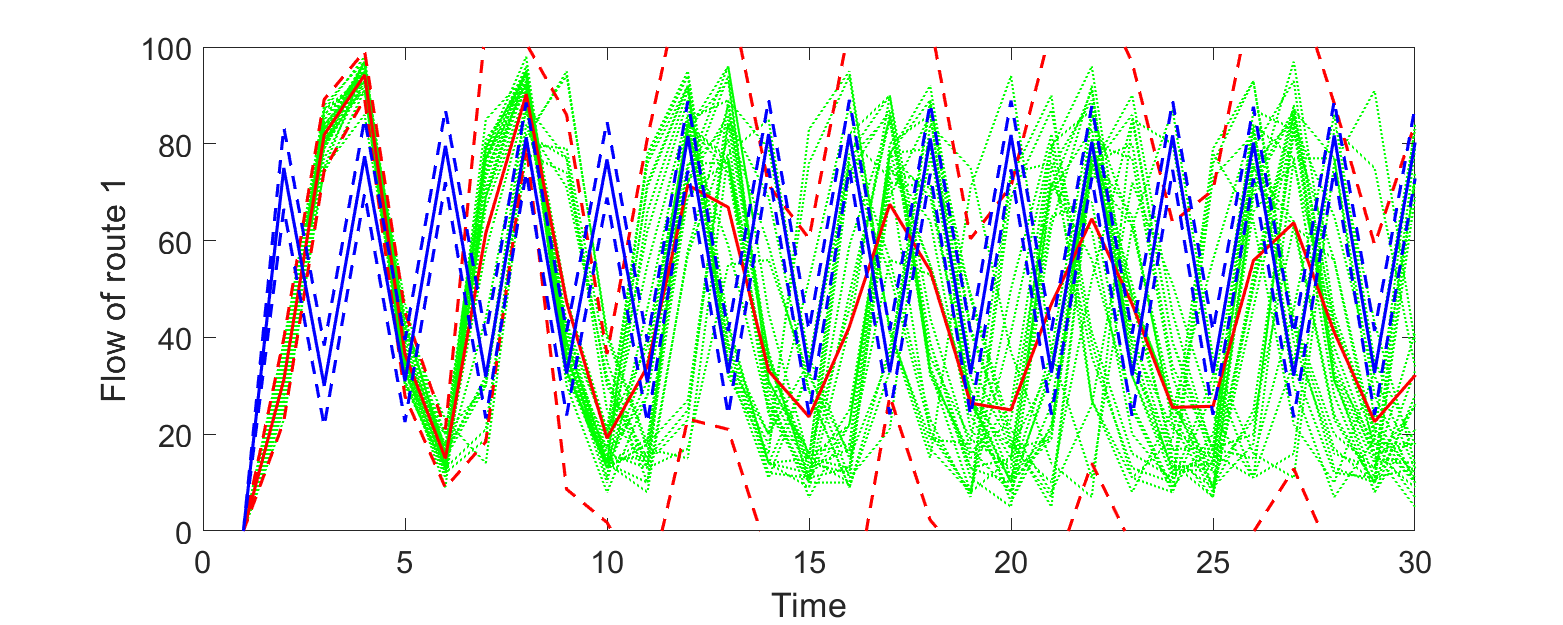}} 
    \caption{Numerical examples of the approximate model. Only the results during the first 30 days are presented for better visualization.}
    \label{fig:NE results}
\end{figure}

Figure \ref{fig:NE results} presents the results. 
As shown by the stability analysis in Section \ref{sec:Stability}, the route changing intensity is very low when $\theta=0.01$ (Figure \ref{fig:NE results}(a)), and thus the route flow difference between two successive days is small enough to guarantee the stability. 
\textcolor{black}{It is also found from Figure \ref{fig:NE results}(a) that the trajectory for a single simulation never converges but oscillates within a certain range.
It is consistent with the observations in the route choice experiments that we introduced in Part I of the research (see Figure 2 in \cite{QI2023103553}). 
On the other hand, the mean trajectory of flow of the route 1 has a smooth transient state (days 1-5) and finally converges to its equilibrium value (after day 5).
Both the mean values and the 95\% confidential levels from the two models are very close, not only in the steady stage (after day 5) but also in the transient state (days 1-5).}

In contrast, if set $\theta=1$, the equilibrium point of the network is never stable, since large $\theta$ means that travelers changes routes intensively even when the cost difference is very small.
Therefore, flow difference between two successive days might be also large, which violates the condition of stability.

The results when $\theta=0.1$ is an interesting 'median' case. 
In this case, the estimation error is large during the initial several days, because the flow changes sharply in the far-from-equilibrium initial states. 
When the system is gradually stable, the estimation error becomes much less for both mean value and  confidential level.

\subsection{Computational Efficiency}\label{sec:eff}

One of the major motivations of proposing the approximate model is to reduce computational burden, and thus we demonstrate the computational efficiency of the approximate model. 
Both the RDAB-SP model and its approximate model are coded using C\# language and run in a personal computer with Windows 10 OS, Intel i7 2.5 GHz CPU, and 8 GB memory.
Table \ref{tab:running time} shows the computational time of the two models. 
It can be seen that the approximate model obviously outperforms the RDAB-SP model in terms of computational efficiency, in particular when the number of travelers is large.

\begin{table}[!htbp]\centering\footnotesize
\setlength{\tabcolsep}{13mm}{
\caption{Comparison of the computational time of the RDAB-SP model and its approximate model (1000 days)}\label{tab:running time}
\begin{tabular}{ccc}
\toprule
Demand &  \multicolumn{2}{c} {Computational time (ms)} \\
\cline{2-3}
 & RDAB-SP Model & Approximate Model  \\
\midrule
1000 &  63 & 0.89\\
10000 &  1649 & 12.9\\
100000 &  16327 & 12.6 \\
\bottomrule
\end{tabular}}
\end{table}

\subsection{Policy Implications}\label{sec:pol}
Both the proposed RDAB-SP model and its approximation are practically valuable for transportation engineering, planning and policy making. 
For example, it is critical for transportation managers to predict daily variations of network flow after a new policy such as adding toll gates or traffic restriction is implemented.
Traditionally, only trends or expected future values of network flow could be estimated by using the deterministic model. 
However, due to the existence of stochastic fluctuations, real values of network flow may diverse largely from expected value, which would negatively affect the performance of policies. 
Now, with the aid of the proposed stochastic models, transportation engineers and planners could qualitatively predict the statistical characteristics of network flow, including mean value, variance, confidential interval, and even probabilistic distribution, which would be very helpful for designing better policies.
Moreover, high computational efficiency of the approximate model could further guarantee the practicability of the proposed methods.






\section{Conclusion}\label{sec:Conclusion}
Limited by the deterministic nature, the differential function-based DTD model can only show the expected or mean evolutionary trajectory of the network flow. 
To understand the driving force behind the random oscillations that were widely observed in experiments, this part of the research proposes a RDAB-SP model based on the behavioral phenomena observed from the DTD route-choice experiments (presented in Part I of the research \citep{QI2023103553}). 
The RDAB-SP model is tested from three indices, i.e., the switching flow, the one-step evolution, and the route flow distribution. 
Moreover, the data source collected from another paper is employed to test the generalization of the proposed model.
The results show that the model can satisfactorily reproduce the random oscillations of a network flow evolution process. 
In addition, an approximate model of the proposed model is proposed to reduce the computational burden, which effectively facilitates the model to be a useful and practical tool for transportation planners and policy makers.

To the best of our knowledge, this paper is the first attempt to explain experimental findings by proposing a stochastic process DTD model. 
It is interesting to find that the seemingly disordered phenomena (random route switching behavior) is actually dominated by simple rules, i.e., independent and probability-based route choice. 
Considering the fact that the RDAB-SP model is directly generated from the deterministic model by replacing the expectation with a distribution, this part of the research further confirms the generality of the basic framework of route choice behavior in Part I, i.e., the discrete choice with route-dependent inertia and preference \citep{QI2023103553}.


Motivated by several interesting observations from a series of route-choice experiments, this research including Parts I and II investigates DTD network flow dynamics based on explicitly modeling travelers' route-choice behavior.
With the proposed analysis, we have achieved better description of DTD route choice behaviors and network flow dynamics from both deterministic and stochastic perspectives.
The key of the research is that travelers' route-choice behaviors are driven by several simple yet interesting rules, and it is important to treat individual traveler's route-choice behavior in a disaggregate way, rather than just concentrating on network-level behaviors such as flow swapping rules.

\textcolor{black} {Some limitations of this study are expected to be addressed in the future. 
The evidence summarized in this paper may be restricted to the networks employed in the experiments, such as linear cost functions, routes with an equal number of edges, single origins, single destinations, and homogeneous users. 
It remains to be seen whether it may be generalized to networks with richer topology, other protocols of play, and a heterogeneous population of network users. 
Further research should explore whether and how the change of any parameter affects the generality of the conclusions and test the performance of the proposed model.}


\section*{Acknowledgement}

The research is funded by National Natural Science Foundation of China (72101085, 71871010) and Laboratory of Computation and Analytics of Complex Management Systems(CACMS) (Tianjin University).


\bibliographystyle{model2-names}
\bibliography{Z.Paper-D2D}

\end{spacing}
\end{document}